\documentclass[amsmath,amssymb,reprint,nofootinbib, onecolumn,longbibliography]{revtex4-1} % One-column

\usepackage{xcolor}
\usepackage{mathtools}
\usepackage{comment}
\usepackage{placeins}

\usepackage{amsmath}
\usepackage{amsfonts}
\usepackage{mathrsfs}
\usepackage{chemformula}

\usepackage{makecell}

\usepackage{soul} % Use \st{} for strikeout, \ul{} for underline

\newcommand{\bs}{\boldsymbol}

\newcommand{\aWien}{\mathfrak{a}}

\newcommand{\kb}{k_{\text{B}}}

\newcommand{\position}{\boldsymbol{x}}
\newcommand{\absposition}{x}
\newcommand{\paramposition}{\tilde{x}}
\newcommand{\gradposition}{\boldsymbol{\nabla}_x}

\newcommand{\spatial}{\boldsymbol{r}}
\newcommand{\absspatial}{r}
\newcommand{\gradspatial}{\boldsymbol{\nabla}_r}
\newcommand{\laplacian}{\boldsymbol{\nabla}_r^2}

\newcommand{\pderiv}[2]{\frac{\partial #1}{\partial #2}}

\newcommand{\graddiscrete}{\boldsymbol{\nabla}_{h}}
\newcommand{\gradstoch}{\widetilde{\boldsymbol{\nabla}}_{h}}

\newcommand{\force}{\boldsymbol{F}}
\newcommand{\forcedensity}{\boldsymbol{f}}

\newcommand{\absforce}{F}

\newcommand{\particlenoise}{\boldsymbol{\mathcal{W}}}
\newcommand{\discreteparticlenoise}{\boldsymbol{W}}
\newcommand{\rfdnoise}{\widehat{\bs W}}
\newcommand{\rfdnoiseabs}{\widehat{ W}}

\newcommand{\SHIBA}{DISCOS }
\newcommand{\SHIBAe}{DISCOS}

\newcommand{\mobility}{\boldsymbol{\mathcal{M}}}
\newcommand{\identity}{\bs{I}}

\newcommand{\radius}{a}

\newcommand{\fluidvel}{\boldsymbol{v}}
\newcommand{\particlevel}{\boldsymbol{V}}

\newcommand{\fluidvelabs}{v}

\newcommand{\fieldnoise}{\boldsymbol{\mathcal{Z}}}
\newcommand{\discretefieldnoise}{\boldsymbol{Z}}
\newcommand{\fulldiscretenoise}{\widehat{\boldsymbol{Z}}}

\newcommand{\hydrospread}{\bs{\mathcal{S}}^\mathrm{hy}}
\newcommand{\hydrointerp}{\bs{\mathcal{J}}^\mathrm{hy}}

\newcommand{\electrospread}{\bs{\mathcal{S}}^\mathrm{es}}
\newcommand{\electrointerp}{\bs{\mathcal{J}}^\mathrm{es}}

\newcommand{\stokes}{\bs{\mathcal{L}}}

\newcommand{\efield}{\bs{E}}
\newcommand{\eparticle}{\bs{E}}

\newcommand{\efieldabs}{{E}}

\newcommand{\chargevec}{\boldsymbol{q}}
\newcommand{\charge}{q}

\newcommand{\hydrokernel}{\delta^\mathrm{hy}}
\newcommand{\electrokernel}{\delta^\mathrm{es}}

\newcommand{\chargedensity}{\varrho}

\newcommand{\radvar}{\tilde{x}}

\newcommand{\cellvol}{\Delta \mathcal{V}}

\newcommand{\ito}{ }
\newcommand{\kinetic}{\diamond}
\newcommand{\stratonovich}{\circ}

\newcommand{\deleted}[1]{}

\newcommand{\MarginPar}[1]   %{{#1}}
{\marginpar{\vskip-\baselineskip % raise the marginpar a bit
\raggedright\tiny\sffamily\hrule\smallskip{\color{red}#1}\par\smallskip\hrule}}

\newcommand{\MarginJBB}[1]   %{{#1}}
{\marginpar{\vskip-\baselineskip % raise the marginpar a bit
\raggedright\tiny\sffamily\hrule\smallskip{\color{black}#1}\par\smallskip\hrule}}

\begin{document}

\title[\SHIBA for Electrolytes]
    %   {A Stochastic Hydrodynamics Immersed Boundary Algorithm for Modeling Electrolytes}
      {A Discrete Ion Stochastic Continuum Overdamped Solvent
      Algorithm for Modeling Electrolytes}

\author{D. R. Ladiges}
\email{DRLadiges@lbl.gov}
\affiliation{Center for Computational Sciences and Engineering, LBNL}
\homepage{https://ccse.lbl.gov}

\author{S. P. Carney}
\affiliation{Department of Mathematics, UT Austin}

\author{A. Nonaka}
\affiliation{Center for Computational Sciences and Engineering, LBNL}

\author{K. Klymko}
\affiliation{Center for Computational Sciences and Engineering, LBNL}

\author{G. C. Moore}
\affiliation{Center for Computational Sciences and Engineering, LBNL}

\author{A. L. Garcia}
\affiliation{Department of Physics and Astronomy, SJSU}

\author{S. R. Natesh}
\affiliation{Courant Institute of Mathematical Sciences, New York University, New York, NY 10012}

\author{A. Donev}
\affiliation{Courant Institute of Mathematical Sciences, New York University, New York, NY 10012}

\author{J. B. Bell}
\affiliation{Center for Computational Sciences and Engineering, LBNL}

\date{\today}

\begin{abstract}
In this paper we develop a methodology for the mesoscale simulation of strong electrolytes. The methodology is an extension of the Fluctuating Immersed Boundary (FIB) approach that treats a solute as discrete Lagrangian particles that interact with Eulerian hydrodynamic and electrostatic fields. In both algorithms the Immersed Boundary (IB) method of Peskin is used for particle-field coupling. Hydrodynamic interactions are taken to be overdamped, with thermal noise incorporated using the fluctuating Stokes equation, including a ``dry diffusion'' Brownian motion to account for scales not resolved by the coarse-grained model of the solvent. Long range electrostatic interactions are computed by solving the Poisson equation, with short range corrections included using 
a novel immersed-boundary variant of the classical Particle-Particle Particle-Mesh (P3M) technique.  Also included is a short range repulsive force based on the Weeks-Chandler-Andersen (WCA) potential. The new methodology is validated by comparison to Debye-H{\"u}ckel theory for ion-ion pair correlation functions, and Debye-H{\"u}ckel-Onsager theory for conductivity, including the Wien effect for strong electric fields. In each case good agreement is observed, provided that 
hydrodynamic interactions at the typical ion-ion separation are resolved by the fluid grid.
\end{abstract}

\maketitle

\section{Introduction}\label{sec:Introduction}

Understanding transport properties in electrolytes is important for the study of
fundamental processes such as electrophoresis, electro-osmosis and electrochemistry that arise in both biological systems \cite{grodzinsky2011fields} and engineered devices \cite{kirby2010micro}
such as catalytic micropumps \cite{verpoorte2002microfluidic,Li2019}, batteries \cite{bachman2016inorganic,Scott2018}, and fuel cells \cite{Andersson_2010,JahnkeETAL_2016,Arsalis2019}.
Many of these phenomena occur at the mesoscale, where the effects of thermal fluctuations must be captured correctly. This has previously motivated the development of fluctuating hydrodynamics \cite{Land2} methods for electrolytic flows \cite{peraud2016low,donev2019fluctuating,donev2019fluctuating2}, in which the system is described using stochastic partial differential equations, which are solved on a grid. Using this approach, thermal fluctuations are captured while avoiding the computational expense inherent in direct molecular simulations \cite{frenkel2001understanding}.
However, one drawback of a purely continuum model is that it is not suitable for mesoscopic systems that contain features that occur on a molecular scale, but which still have an important effect on the overall dynamics.
In the context of electrolytes the electrical double layer \cite{bockris1965structure} effect is an important example of such phenomena.\footnote{For example, the thickness of the double layer is roughly one Debye length, $\lambda_D$, and for a molar concentration of 1.0M there is only one ion per $(3\lambda_D)^3$.}

In order to capture this behavior while avoiding the computational expense of molecular dynamics, an approach is required that combines a particle model for the solutes with a coarse-grained model of the solvent. Currently, the most well established numerical method that uses this approach is Brownian dynamics (BD) \cite{fixman1978simulation}. The key feature of this technique is the use of Green's functions to represent particles immersed in an implicit solvent. These Green's functions describe the effects of single particles, which can be joined together in groups to form complex molecules such as polymers, or even structures such as flexible membranes. While this approach works well in many cases, it has a range of shortcomings. First, in a basic implementation, the calculation of inter-particle hydrodynamic forces has computational complexity $O(N^2)$, where $N$ is the number of particles being simulated. Further, the calculation of the thermal noise applied to each particle scales as $O(N^3)$ (from the Cholesky factorization of the mobility matrix), making simulation of large systems infeasible. 
Additionally, the application of boundary conditions is complicated to implement in all but the simplest cases.

More recently, an approach referred to as General Geometry Ewald-like Method (GGEM) \cite{hernandez2007fast,zhao2017parallel} has been proposed. Similar in principle to the Particle-Particle Particle-Mesh (P3M) 
method \cite{Hockney:1988:CSU:62815} used in electrostatics, short range hydrodynamic interactions are computed using Green's functions, and long range interactions are computed using a grid-based discretization of the relevant PDE, in this case Stokes' equation.
With suitable preconditioning of the Stokes solver and choice of the splitting parameter, this achieves near linear scaling for deterministic hydrodynamics by limiting the calculation of pairwise interactions to small subsets of particles. In principle, GGEM allows for handling of nontrivial boundary conditions, although some approximations are introduced \cite{hernandez2007fast,zhao2017parallel} to handle the boundaries in the near field;\footnote{In particular, in GGEM, the boundary condition imposed in the Stokes solver involves the near field, which varies on scales that \emph{cannot} be resolved by the grid. A further point-particle approximation is made in the boundary conditions to keep the mobility positive semi-definite \cite{hernandez2007fast,zhao2017parallel}.} see Ref.~\citenum{SpectralEwald_Wall} for exact handling of a bottom wall geometry with Ewald splitting. Importantly, the handling of Brownian motion in GGEM is based on methods introduced in the 1970s by Fixman, and these methods increase the computational cost over deterministic simulations many-fold \cite{SpectralRPY}. First, generating the Brownian displacements uses an iterative method, which ideally employs preconditioners \cite{SquareRootKrylov,SquareRootPreconditioning} in order to control the number of iterations. Second, when boundaries are present, generating the correct stochastic drift required to achieve discrete fluctuation-dissipation balance relies on a method of Fixman that requires generating the action of the square root of the resistance matrix, which itself requires a slowly-converging iterative method.

A related approach that addresses these issues is the Fluctuating Immersed Boundary (FIB) \cite{delong2014brownian} method. This technique uses Immersed Boundary (IB) \cite{peskin2002} kernels instead of Green's functions to couple particles to an explicitly simulated solvent. Importantly, whereas both the original IB and GGEM methods use standard deterministic hydrodynamics to represent the solvent, in FIB the solvent is simulated using an incompressible fluctuating hydrodynamics methodology \cite{nonaka2015low}. Using this approach, thermal noise arises directly from the fluid, avoiding the complex per particle calculations required by GGEM and other Brownian dynamics methods. In FIB, no iterative methods are required to generate either the Brownian increments or the stochastic drift (rather, they are computed together with the deterministic displacements or generated by a simple ``random finite difference'' midpoint scheme inspired by but different from Fixman's), and discrete fluctuation-dissipation balance is ensured exactly by construction even in the presence of boundaries. Linear time complexity, in both domain size and particle number, is achieved for the calculation of hydrodynamic interactions and Brownian motion under all conditions. Additionally, the application of confining boundary conditions, such as no-slip walls, is trivial when using FIB. We note a similar approach is taken in the Stochastic Eulerian Lagrangian Method (SELM), described in Ref.~\citenum{atzberger2011}; some comments regarding the relationship between FIB and SELM are given in Ref.~\citenum{delong2014brownian}. 

Recently, the FIB approach to Brownian motion has been combined with the Ewald splitting used in GGEM for periodic systems in the Positively Split Ewald (PSE) method \cite{SpectralRPY}; we note that the Hasimoto splitting used in PSE is the Fourier space equivalent of the real-space splitting used in GGEM \cite{SE_FreeSpaceStokes}. PSE is a linear-scaling, spectrally-accurate, and grid-independent method that has been shown \cite{SpectralRPY} to be an order of magnitude faster than traditional approaches based on the techniques developed by Fixman. However, including boundaries in spectral-Ewald methods relying on Fourier transforms such as PSE is nontrivial and has not, to our knowledge, been done yet even for deterministic simulations.

In this paper, we extend the FIB method to the simulation of electrolytic flows. Using this approach, individual charged ions are described using immersed boundary kernels, and the solvent is treated using fluctuating hydrodynamics. In doing so we incorporate two new contributions to the FIB methodology: (i) a ``dry diffusion'' process which implements a coarse-grained model for small ions (e.g., $\mathrm{Na}^+$),
based on the approach derived in Ref.~\citenum{espanol2015coupling}, and (ii) the incorporation of electrostatic effects with an immersed-boundary P3M implementation for efficiently and accurately computing the electric field. We refer to this approach as Discrete Ion Stochastic Continuum Overdamped Solvent (DISCOS). 

The layout of this paper is as follows. In Sec.~\ref{sec:BrownFIB}, we summarize the Brownian dynamics approach and describe its relationship to the FIB method. In Sec.~\ref{sec:diffusion} we discuss the dry diffusion process.  Sec.~\ref{sec:IonicForces} contains a description of electrostatic and close range forces (with additional detail given in appendices \ref{sec:IB-P3M} and \ref{sec:IB-P3M-Tables}), 
while Sec.~\ref{sec:NumericalMethods} details the numerical methodology used to implement the \SHIBA algorithm. In Sec.~\ref{sec:Results} the \SHIBA method is tested by comparison to theoretical results for the radial distribution function and electrical conductivity. 
This includes an analysis of the effect of the dry diffusion approach described in Sec.~\ref{sec:diffusion}. Finally, some concluding remarks are given in Sec.~\ref{sec:Conclusion}.
Note that in this article, to fully detail and test the novel aspects of the method without introducing additional complications, we describe the application of \SHIBA to unbounded/periodic systems. The inclusion of boundaries will be the subject of a future publication, however we have included a brief discussion in Appendix~\ref{sec:boundaries}.

\section{Stochastic Hydrodynamics}
\label{sec:BrownFIB}

This section formulates stochastic hydrodynamics for Brownian particles, and Sec. \ref{sec:IonicForces} describes the additional intermolecular forces present when these particles are ions.
First we briefly review the Brownian dynamics and fluctuating immersed boundary approaches to simulating particle-fluid systems; more detailed descriptions can be found in Refs.~\citenum{fixman1978simulation,delong2014brownian,Stoch_DNA,hernandez2007fast} and \citenum{confine_DNA}. 
Both approaches assume an infinite Schmidt number,
\begin{align}
\mathrm{Sc} = \frac{\displaystyle \eta}{\displaystyle \rho D}\rightarrow \infty,
\end{align}
where $D$ is the diffusion coefficient of a particle, and $\eta$ and $\rho$ are the viscosity and density of the fluid, respectively.  In this asymptotic regime, the diffusion time of the particles is large compared to the relaxation time of the fluid, and the flow can therefore be treated as quasi-steady. A detailed discussion of the relationship between particle diffusion and Schmidt number is given in Ref.~\citenum{balboa2013stokes}, and the validity of this approximation in the context of electrolytes is discussed in Sec.~\ref{sec:Results}. 

The FIB method was originally intended as a fast algorithm for performing Brownian dynamics with hydrodynamic interactions for colloidal suspensions; the reader can consult Refs.~\citenum{SpectralRPY} and  \citenum{SE_Multiblob_SD} for related state-of-the-art spectrally-accurate methods for periodic suspensions. Here we adapt this methodology to ionic solutions. This necessitates some important changes to account for the fact that ions are much smaller than typical colloids, as we describe in detail in Sections \ref{sec:fib} and \ref{sec:diffusion}.

\subsection{Brownian dynamics with hydrodynamic interactions}\label{sec:brownian}
We consider a solute (ions) represented as $N$ particles with positions $\position(t) = \{\position_1,\cdots,\position_i,\cdots,\position_N\}$. The equation of motion for the particles is
\begin{align}
\frac{d\position}{dt} &= \mobility \force  + \sqrt{2 \kb T}\mobility^{1/2}\kinetic\particlenoise
\label{eq:brownian1}\\
&= \mobility \force  + \sqrt{2 \kb T}\mobility \stratonovich \mobility^{-1/2}\ito\particlenoise.
\label{eq:brownian3}
\end{align}
Here, $\force(\position,t) =  \left\{\force_1,\cdots,\force_i,\cdots,\force_N\right\}$ are the net forces acting on the particles (in this article they consist of short-ranged intermolecular as well as long-ranged electrostatic forces),
 $\particlenoise(t) =  \left\{\particlenoise_1,\cdots,\particlenoise_i,\cdots,\particlenoise_M\right\}$ are independent Gaussian white noise processes,\footnote{Note that it is not required that $M=N$.} $\kb$ is Boltzmann's constant, and $T$ is the temperature. The stochastic product symbol $\kinetic$ indicates the kinetic interpretation \cite{klimontovich1990ito,hutter1998fluctuation} of the stochastic integral, which can be viewed a mixed Stratonovich-It\^{o} interpretation, with $\stratonovich$ denoting the Stratonovich and the absence of a symbol denoting the It\^{o} stochastic product. The symmetric positive-definite mobility matrix $\mobility(\position)$ encodes the hydrodynamic interactions between particles.
Here, we define $\mobility^{1/2}$ and $\mobility^{-1/2}$ to be (not necessarily square\footnote{One may view the FIB/FHD approach as corresponding to a non-square decomposition of the mobility matrix. In this case the number of noise terms will be proportional to the number of grid points, rather than to the number of particles; by construction the noise will have the correct covariance.}) matrices such that
\begin{align}
\mobility&=\mobility^{1/2} \left(\mobility^{1/2}\right)^\star,\nonumber\\
\mobility^{-1}&=\mobility^{-1/2} \left(\mobility^{-1/2}\right)^\star,
\label{eq:spd}
\end{align}
where the $\star$ indicates $L_2$ adjoint (conjugate-transpose). 
When the kinetic interpretation of the stochastic integral is applied this property ensures fluctuation-dissipation balance is obeyed, i.e., that the equilibrium dynamics are time reversible with respect to the Gibbs--Boltzmann distribution. 
When we recast Eq.~(\ref{eq:brownian1}) as a system of It\^o stochastic differential equations for numerical simulation,\
it is necessary to include a stochastic drift term, yielding
\begin{equation}
\frac{d\position}{dt} = \mobility \force  +  \kb T \gradposition \cdot \mobility + \sqrt{2 \kb T}\mobility^{1/2}\ito\particlenoise,
\label{eq:brownian2}
\end{equation}
where $\gradposition \cdot$ is the divergence with respect to the particle position variables.

In Brownian dynamics the main considerations in the numerical integration of Eq.~(\ref{eq:brownian2}) are the construction and fast application of $\mobility$, and the subsequent application of $\mobility^{1/2}$ and calculation of \mbox{$\gradposition \cdot \mobility$}. The action of a point force at the origin on the fluid is given by the Green's
function solution to the steady Stokes' equation, the Oseen tensor
\begin{align}
\bs{\mathcal{O}}(\spatial)= \frac{1}{8 \pi \eta \absspatial }\left(\identity+\frac{\displaystyle \spatial \otimes \spatial }{\displaystyle \absspatial^2} \right).
\end{align} 
Here, $\spatial$~is the position vector with $\absspatial=|\spatial|$ and $\bs{I}$ is the identity tensor. This can be modified to include finite size \cite{kekre2010comparison} and close range corrections, giving the Rotne-Prager-Yamakawa (RPY) tensor \cite{rotne1969variational},
\begin{align}
\bs{\mathcal{R}}(\spatial;a)= 
\frac{1}{6 \pi \eta \radius }
\begin{cases} 
     C_1 \identity+C_2 \frac{\displaystyle \spatial \otimes \spatial }{\displaystyle \absspatial^2} \vspace{2mm}& \absspatial> 2 \radius \\
      C_3 \identity+C_4 \frac{\displaystyle \spatial \otimes \spatial }{\displaystyle \absspatial^2} &  \absspatial \leq 2 \radius
   \end{cases},\label{eq:rpy}
\end{align}
with
\begin{align}
     &C_1 = \frac{\displaystyle 3 \radius }{\displaystyle 4 \absspatial} + \frac{\displaystyle \radius^3 }{\displaystyle 2 \absspatial^3} , \hspace{3mm} C_2= \frac{\displaystyle 3\radius }{\displaystyle 4 \absspatial} - \frac{\displaystyle 3 \radius^3 }{\displaystyle 2 \absspatial^3},\nonumber \\
     &C_3 =1-\frac{\displaystyle 9 \absspatial }{\displaystyle 32 \radius} , \hspace{5mm}  C_4= \frac{\displaystyle 3 \absspatial }{\displaystyle 32 \radius},
\end{align}
where we have assumed all particles have the same hydrodynamic radius $\radius$. The $3\times3$ block of the mobility matrix yielding the velocity of particle $i$ due to a force applied to 
 particle $j$ is then 
\begin{align}
\mobility_{ij} = \bs{\mathcal{R}}(\position_{ij};\radius),\label{eq:submob}
\end{align}
where $\position_{ij} = \position_{i} - \position_{j}$. The complete mobility matrix is assembled by computing Eq.~(\ref{eq:submob}) for all particle pairs.
We note that $\mobility$ is symmetric positive-definite by construction.
Having obtained $\mobility$, the matrix $\mobility^{1/2}$ can be calculated exactly using Cholesky decomposition or applied using a more rapid iterative method of approximation \cite{fixman1986construction,SquareRootKrylov}.

Rather than directly calculating the divergence of the mobility matrix to obtain the stochastic drift, the drift is traditionally included via the Fixman \cite{fixman1978simulation, grassia1995computer} midpoint time stepping algorithm. This can also be viewed as a direct discretization of the kinetic version of the stochastic integral \cite{hutter1998fluctuation}
or, equivalently, a direct discretization of the mixed-integral in Eq.~\eqref{eq:brownian3}: 
\begin{align}
\position^{n+1/2,\star} &= \position^n + \frac{\Delta t}{2}\mobility(\position^n)\force^n + \sqrt{\frac{\Delta t \kb T}{2}}(\mobility^n)^{1/2}\discreteparticlenoise^n ,\nonumber\\
\position^{n+1} &= \position^n+\Delta t \mobility(\position^{n+1/2,\star})\left(
\force^{n+1/2,\star} 
+ \sqrt{\frac{2 \kb T}{\Delta t}} \left(\mobility^{(n)}\right)^{-1/2}\discreteparticlenoise^n
\right).\label{fixman}
\end{align}
Here, $\Delta t$ is the discrete time step and $\discreteparticlenoise$ is a vector of independent Gaussian random variables. The superscripts $n$ and $n+1$ denote values at consecutive time steps, with $n+1/2,\star$ indicating a midpoint update.

\subsection{Fluctuating Hydrodynamics Formulation}\label{sec:fib}
As discussed in Sec.~\ref{sec:Introduction}, the FIB method differs from Brownian dynamics principally in that it explicitly simulates a coarse grained model of the solvent~\cite{espanol2015coupling} on an Eulerian grid.
The solvent is taken to be isothermal and incompressible, and is therefore modeled by the fluctuating Stokes equations \cite{donev_mesoscopic_diffusion:2014,espanol2015coupling},
\begin{subequations}
\begin{align}
\rho \pderiv{\fluidvel}{t}  + \bs \nabla_r p - \eta \laplacian \fluidvel =&\, \forcedensity + \sqrt{2 \kb T \eta}\, \gradspatial \cdot \fieldnoise,\label{eq:stokesA}\\
\bs{\nabla}_r \cdot \bs v =&\, 0,
\end{align}\label{eq:stokes}%
\end{subequations}
where $\bs \nabla_r$ is the gradient operator with respect to $\spatial$, $\fluidvel(\spatial,t)$ is the fluid velocity, $p(\spatial,t)$ is the pressure, and 
 $\forcedensity(\spatial,t)$ is a force density applied to the fluid; this is the mechanism by which the immersed Brownian particles interact with the solvent. Finally, $\fieldnoise(\spatial,t)$ is a random, symmetric, spatio-temporal Gaussian tensor field (the stochastic stress tensor) whose components are white in space and time, i.e., they have mean zero and covariances 
\begin{align}
\langle \mathcal{Z}_{ij}(\spatial,t)\mathcal{Z}_{kl}(\spatial',t')\rangle &= (\delta_{ik}\delta_{jl} + \delta_{il}\delta_{jk})\delta(t-t')\delta(\spatial-\spatial') \label{eq:covar}\\
&\equiv \mathbf{\Sigma} \; \delta(t-t')\delta(\spatial-\spatial') . \nonumber
\end{align}
Here, $\delta_{ij}$ is the Kronecker delta function, $\delta$ is the Dirac delta function, $\mathbf{\Sigma}$ is the pointwise covariance of the noise, and $\langle \cdots \rangle$ denotes an ensemble average. This time-dependent system is a set of linear, stochastic partial differential equations with additive noise, and it therefore has a well-defined mathematical interpretation \cite{da_prato}. Furthermore, it can be shown to satisfy fluctuation-dissipation balance \cite{atzberger2011,espanol2015coupling}.

In Refs.~\citenum{donev_mesoscopic_diffusion:2014,delong2014brownian,atzberger2011} the diffusing particles are coupled directly to the continuum equations, Eq.~\eqref{eq:stokes}. However, as discussed in detail in Refs.~\citenum{espanol2015coupling,Donev_10}, fluctuating hydrodynamics (FHD) is a coarse-grained description that only has a clear interpretation once the conserved fluid quantities (mass, momentum, energy) are spatially coarse-grained on a discrete grid. Each hydrodynamic cell should contain a sufficiently large number of solvent molecules for the coarse-grained description to be justified. Since our particles are (solvated) ions of size comparable to the solvent molecules, this implies that the fluid grid size must be larger than a diffusing ``nano-particle''. This is exactly the case analyzed in Ref.~\citenum{espanol2015coupling} using the theory of coarse-graining, where the discrete FHD equations are derived by coupling a single nano-particle with compressible, isothermal FHD. We base our discrete equations on this derivation, and take advantage of the flexibility of the framework to improve numerical properties, such as grid-independence of physical observables and preservation of translational invariance.

One key conclusion of the analysis in Ref.~\citenum{espanol2015coupling} is that the total or effective diffusion coefficient of an immersed particle consists of two pieces. The first comes from the random advection of the particle by the coarse-grained stochastic fluid velocity; we will refer to this process as ``wet diffusion'', with the diffusion coefficient $D^{\rm wet}$. The second comes from the additional random motion of the particle relative to the coarse-grained fluid velocity; we will term this ``dry diffusion'', with the coefficient $D^{\rm dry}$. The sum of these two terms gives the total diffusion coefficient of the particle, $D^{\rm tot}$.
One can formally write a Green-Kubo formula expressing the dry diffusion coefficient as the integral of the autocorrelation function of the particle velocity relative to the coarse-grained fluid velocity \cite{espanol2015coupling}; we discuss how we set the value of the dry diffusion coefficient in Sec.~\ref{sec:diffusion}.

\subsection{Steady, overdamped limit}

For high Schmidt numbers the fluid relaxation time is fast on the time scale of particle diffusion, so
the time-dependent system Eq.~\eqref{eq:stokes} can
be accurately approximated by
a (quasi\nobreakdash-)steady 
Stokes system. This limiting behavior, referred to as
the overdamped limit, can be derived by a rigorous analysis of the limit as $\mathrm{Sc}\to \infty$; this is shown in Appendix A of Ref.~\citenum{donev_mesoscopic_diffusion:2014}.
In the overdamped limit, the dynamic variables are just the positions of the immersed particles, and they follow an equation that is identical in structure to the overdamped Langevin equation, Eq.~\eqref{eq:brownian1}.

Taking the subscript $h$ to indicate a discrete operator, in the overdamped limit we take the discrete fluid velocity and pressure to satisfy the steady-state form of Eq.~(\ref{eq:stokes}) given by
\begin{subequations}
\begin{align}
\graddiscrete p - \eta \graddiscrete^2 \fluidvel =&\, \forcedensity + \sqrt{\frac{2 \kb T \eta}{\cellvol }}\, \graddiscrete\cdot \discretefieldnoise,\label{stokes_mollified}\\
\graddiscrete \cdot  \bs v =&\, 0,
\end{align}\label{eq:disc_stokes}%
\end{subequations}
where
 $\discretefieldnoise(\spatial_h, t)$ is a finite dimensional collection of white noise processes representing the spatial discretization of $\fieldnoise$ on a regular grid with positions $\spatial_h$, and $\cellvol$ is the cell volume.
Preserving the fluctuation-dissipation property then requires that the discrete divergence be the negative of the $L_2$ adjoint of the discrete gradient, $\graddiscrete^\star=-\graddiscrete \cdot$, and that
\begin{equation}\label{eq:compatibility_on_noise}
    \nabla^2_h = \graddiscrete \cdot \mathbf{\Sigma}
    \graddiscrete = -\graddiscrete ^\star \mathbf{\Sigma} \graddiscrete.
\end{equation}
The spatial discretization scheme describing the relationship between the continuum operators/fields of Eq.~(\ref{eq:stokes}) with discrete versions in Eq.~(\ref{eq:disc_stokes}) is given in Sec.~\ref{sec:NumericalMethods}; it is designed to ensure that the above properties hold. Note also that comparing Eqs.~(\ref{eq:stokes}) -- (\ref{eq:disc_stokes}) indicates the variance of the noise term $\discretefieldnoise$ is $1/\cellvol$. Intuitively this follows from the fact that a larger cell volume represents a coarse graining over a larger number of solvent particles; this is discussed further in Ref.~\citenum{espanol2015coupling}.

\subsection{Particle-fluid coupling}
In the FIB method the spatial extent of each particle is defined by a compact kernel function $\hydrokernel(\spatial)$, where the superscript ``$\mathrm{hy}$'' indicates that this kernel applies to hydrodynamic fields (the analogous electrostatic kernel will be denoted $\electrokernel$).
The kernel is used to interpolate the local fluid velocity to the particles' locations $\position_i$, and to define the region over which force on the particle is transmitted (spread) to the fluid; the functional form of the kernel is discussed in Section~\ref{sec:peskin}. To perform these operations, we define discrete interpolation and spreading operators, $\hydrointerp_h(\position_i)$ and $\hydrospread_h(\position_i)$, such that
\begin{align}
\particlevel_i  =&\,
\hydrointerp_h(\position_i)\bs{v} = \int \hydrokernel(\position_i - \spatial_h)\bs{v}(\spatial_h) d\spatial_h,\label{kernel1}\\
\forcedensity^\text{s}(\spatial_h) =&\, \left(\hydrospread_h(\position_i)\force\right)(\spatial_h) = \sum_{i=1}^N \hydrokernel(\position_i-\spatial_h)\force_i,\label{kernel2}
\end{align}
where the hydrodynamic (advective) part of the particle velocity, i.e., the velocity of the fluid at the location of the particles, is denoted by $\bs{V} = \left\{\particlevel_1,\cdots,\particlevel_i,\cdots,\particlevel_N\right\}$. The force density spread to the fluid from the particles is given by $\forcedensity^\text{s}$, and as above, we use the subscript $h$ to indicate discrete operators.
Note that the integral in Eq.~\eqref{kernel1} is just a short-hand notation for a weighted sum over the grid points; the exact form of the discretization is discussed in Sec.~\ref{sec:peskin}.
  
It is important to note that this coupling of the particles to the coarse-grained fluid, taken from immersed boundary methodology, is different from (though closely related to) that used in the theory of coarse-graining given in Ref.~\citenum{espanol2015coupling}. The FIB formulation is advantageous because it is simpler to implement, achieves much better translational invariance in practical numerical implementations, and allows flexibility in choosing $\hydrokernel$.

In addition to advection by the velocity given by Eq.~\eqref{kernel1}, particles move by the dry diffusion process, giving the complete overdamped equations of motion for the \SHIBA method (see also Appendix B.1 in Ref.~\citenum{ISIBM}),
\begin{equation}
\frac{d\position_i}{dt} = \underbracket[0.4pt][2pt]{\bs{V}_i \vphantom{ \frac{D_i^{\rm dry}}{\kb T}\force_i }       }_{\text{wet}} + \underbracket[0.4pt][2pt]{\frac{ D_i^{\rm dry}}{\kb T}\force_i + \sqrt{2D_i^{\rm dry}}\ito\particlenoise^{\rm dry}_i}_{\text{dry}},
\label{eq:ions}
\end{equation}
where $\particlenoise^{\rm dry}_i(t)$ is another independent Gaussian white noise process. Note that here we have written the equation of motion for a single particle, since $D_i^{\rm dry}$ depends on the particle's species. 

We define the discrete Stokes operator $\stokes_h$ such that the solution of Eq.~\eqref{eq:disc_stokes} is
\begin{equation}
\fluidvel = \stokes^{-1}_h \left(\forcedensity + \sqrt{\frac{2 \kb T \eta}{\cellvol}}\, \graddiscrete\cdot   \discretefieldnoise \right),
\end{equation}
so that the particles' hydrodynamic velocity is
\begin{align}
\particlevel_i =& \hydrointerp_h(\position_i)\stokes_h^{-1}\left(\hydrospread_h(\position_i)\force(\position)+ \sqrt{\frac{2 \kb T \eta}{\Delta \mathcal{V}}}\, \graddiscrete\cdot \ito \discretefieldnoise\right).\label{eq:ions2}
\end{align}
This allows us to relate the \SHIBA method to Brownian dynamics (Eqs.~(\ref{eq:brownian1}) and (\ref{eq:brownian2})) by observing tha
\begin{equation}
 \mobility = \text{Diag}\left\{\frac{\bs{D}^{\rm dry}}{\kb T}\right\} + \hydrointerp_h\stokes^{-1}_h \hydrospread_h,
 \label{eq:mob_SHIBA}
\end{equation}
where $\bs{D}^{\rm dry}=\left\{ D_1^{\rm dry},\cdots,D_i^{\rm dry},\cdots,D_N^{\rm dry} \right\}$, and $\text{Diag}\{ \bs{X} \}$ denotes a diagonal matrix with the values of $\bs{X}$ on the diagonal. The Brownian velocity is expressed as
\begin{align}
 \mobility^{1/2} \particlenoise \equiv  \text{Diag}\left\{\sqrt{\frac{ \bs{D}^{\rm dry}}{\kb T}}\right\}&\particlenoise^{\rm dry}_i + \sqrt{\frac{\eta}{\Delta \mathcal{V}}} \hydrointerp_h\stokes^{-1}_h  \graddiscrete\cdot\discretefieldnoise,
 \label{eq:sqrtmob_SHIBA}
\end{align}
where $\sqrt{\bs{D}^{\rm dry}} = \left\{ \sqrt{D_1^{\rm dry}},\cdots,\sqrt{D_i^{\rm dry}},\cdots,\sqrt{D_N^{\rm dry}} \right\}$.
In Ref.~\citenum{delong2014brownian} (see also Appendix B of Ref.~\citenum{ISIBM}) it is demonstrated that this combination satisfies fluctuation-dissipation balance, provided Eq.~\eqref{eq:compatibility_on_noise} holds, and the spreading and interpolation are related by the adjoint property $\left(\hydrointerp_h\right)^\star = {\Delta \mathcal{V}}\hydrospread_h$, which is assured by the immersed-boundary method.

What is now needed is a way to evaluate the stochastic drift. Direct implementation of the Fixman method (Eq.~(\ref{fixman})) is complicated by the fact that it requires the inverse of the mobility matrix, which is difficult to obtain in the context of FIB. As demonstrated in Ref.~\citenum{delong2014brownian}, the divergence of the mobility can be evaluated numerically using a finite difference method. However this would require multiple evaluations of $\mobility$ per step, which requires multiple applications of $\stokes_h^{-1}$, i.e., multiple solves of Stokes equation. This can be avoided by re-writing the stochastic drift using the chain rule
\begin{align}
& \gradposition \cdot \mobility = \hydrointerp_h\stokes^{-1}_h \gradposition\cdot \hydrospread_h+ \left( \gradposition \hydrointerp_h\right):\left(\stokes^{-1}_h \hydrospread_h\right)\label{eq:difbypart}
\end{align}
where, in summation notation, the double dot product is defined as
\begin{align}
\left\{\left( \gradposition \hydrointerp_h\right):\left(\stokes^{-1}_h \hydrospread_h\right)\right\}_{ij}\hspace{-2mm} =\hspace{-0.5mm}
(\partial_j (\mathcal{J}^\mathrm{hy}_h)_{ik})(\mathcal{L}_h)^{-1}_{kl} (\mathcal{S}^\mathrm{hy}_h)_{lj}.
\end{align}
Defining the thermal forcing
\begin{equation}
\forcedensity^{\rm th} = \kb T  \gradposition\cdot \hydrospread_h,\label{eqn:thermforce}
\end{equation}
the first term on the right hand side of Eq.~(\ref{eq:difbypart}) is then accounted for by including $\bs f^{\text{th}}$ as part of the force density in 
Eq.~\eqref{eq:disc_stokes},
\begin{equation}
\forcedensity = \forcedensity^{\rm s} + \forcedensity^{\rm th}.
\label{eq:total_f_Stokes}
\end{equation}
The divergence of the spreading operator can be evaluated with a single application of $\stokes^{-1}_h$; the details are discussed in Sec.~\ref{sec:NumericalMethods}.
The second part of the stochastic drift (corresponding to the second term on the right hand side of Eq.~(\ref{eq:difbypart})) is obtained by a midpoint temporal integrator described in Sec. \ref{subsec:TemporalAlgorithm}, similar in spirit to how the Fixman midpoint scheme (Eq.~\eqref{fixman}) obtains the total stochastic drift. Some alternative approaches to evaluating the stochastic drift correction are discussed in Refs.~\citenum{delmotte2015simulating} and \citenum{de2016finite}.

\subsection{Total, wet, and dry diffusion}\label{sec:diffusion}

At the end of Sec.~\ref{sec:fib} we describe two diffusion processes: wet diffusion, which arises from fluctuations in the coarse grained hydrodynamic solution, and dry diffusion, which represents additional Brownian motion that is not resolved by the coarse graining process. Here we explain how we set the dry diffusion coefficient.

Consider a single freely diffusing isolated spherical nanoparticle suspended in a quiescent fluid. The particle will perform a standard Brownian motion with a ``total'' diffusion coefficient $D^{\rm tot}$.
From the Stokes-Einstein relation, the total diffusion coefficient for a sphere with radius $a_t$ is
\begin{equation}
D^{\rm tot} = \frac{\kb T}{\varsigma \eta \radius_t},\label{eq:diffTotal}
\end{equation}
where $\varsigma$ is a constant depending on the boundary condition on the particle; here we take $\varsigma = 6\pi$, corresponding to a no slip boundary condition. Note that as long as $D^{\rm tot}$ is held fixed the relative value of $\varsigma$ and $\radius_t$ will have no bearing on the dynamics. Note also that the radius $a$ used in Eq.~(\ref{eq:rpy}) corresponds to $a_t$.

As described in Sec.~\ref{sec:fib}, when simulating such a particle using the FIB method the fluctuations in the hydrodynamic velocity will yield a wet diffusion process with coefficient $D^{\text{wet}}$. Because these fluctuations are transmitted to the particle via the interpolation operator, $\hydrointerp_h$, the effective hydrodynamic radius of the particle will depend on the form of the kernel function $\hydrokernel$; we designate this wet radius $a_w$, where
\begin{equation}
D^{\rm wet} = \frac{\kb T}{\varsigma \eta \radius_w}.
\end{equation}
The value of $a_w$ is determined by the grid used to solve Eq.~(\ref{eq:disc_stokes}) -- coarsening this grid will result in an increase of $a_w$ and a reduction in the diffusion experienced by the particle.

If the discretization is such that $a_w > a_t$, then the additional dry diffusion process to be added is such that
\begin{equation}
D^{\rm dry} = \frac{\kb T}{\varsigma \eta \radius_d},
\end{equation}
where the dry radius $\radius_d$ is
\begin{align}
    \radius_d = \frac{\radius_w \radius_t}{\radius_w - \radius_t},\label{radsum}
\end{align}
giving
\begin{equation}
D^{\rm tot} = D^{\rm wet} + D^{\rm dry}.\label{diffSum}
\end{equation}
In order for the dry diffusion to remain positive, the total radius $a_t$ must be smaller than the
wet radius $a_w$.
In practice, $D^{\rm wet}$ is set by the mesh spacing and the choice of kernel. We then specify $D^{\rm dry}$ so that each species diffuses with the desired total diffusion coefficient.
Using this approach, the diagonal terms in an under-resolved mobility matrix are corrected with the dry diffusion terms; see Eq.~\eqref{eq:mob_SHIBA}. This has the effect of greatly improving computational speed, at the cost of neglecting short range hydrodynamic contributions. For electrolytes, in many cases we expect these contributions to be small compared to the electrostatic effects; this is discussed further in Section~\ref{subsec:Conductivity}.

The above procedure was suggested, though not implemented, in Ref.~\citenum{espanol2015coupling}, where the total, wet, and dry diffusion coefficients are referred to as the renormalized, enhancement, and bare diffusion coefficients, respectively. A similar approach is also used for inertial systems in Refs.~\citenum{ahlrichs1999simulation,ahlrichs1998lattice,ISIBM}. 
Note that here we have taken the dry diffusion coefficient to be isotropic and spatially homogeneous, which is true only in the case of unbounded/periodic domains. In this case, no stochastic drift correction is necessary for the dry component of Eq.~(\ref{eq:ions}). For other boundary conditions the dry diffusion % if the 
coefficients can be spatially dependent, and a corresponding stochastic drift term would have to be included.

\section{Ionic Forces}\label{sec:IonicForces}
As discussed in Sec.~\ref{sec:Introduction}, we are considering long range electrostatic and short range repulsive forces acting on the ions. For a given ion $i$, the total force is given by
\begin{align}
\force_i = \force^{\mathrm{E}}_{i} + \sum_{j\in\Omega_i^{\mathrm{R}}}\force^{\mathrm{R}}_{ij} + \force^{\mathrm{ext}}_{i}, \label{eq:totalforce}
\end{align}
where $\force^\mathrm{E}_{i}$ is the electrostatic force, $\force^\mathrm{ext}_{i}$ indicates forces due to an applied field (e.g., gravity, an external electric field from a source outside the simulation domain), $\force^\mathrm{R}_{ij}$ is the short range repulsive force between particles $i$ and $j$, and $\Omega_i^{\mathrm{R}}$ indicates all particles within a given range of the  $i^\mathrm{th}$ particle.

\subsection{\label{subsec:Electrostatics}Electrostatic Forces}
First, we describe our immersed-boundary variant of the classical 
Particle-Particle, Particle-Mesh (P3M) approach \cite{Hockney:1988:CSU:62815, frenkel2001understanding} in detail, since we are not aware of prior work that implements an IB-P3M algorithm. We employ an approach that is essentially linear in system size to evaluate electrostatic interactions, while recovering a point charge representation of the ions which cannot be captured using purely grid based methods. For periodic systems, the current state-of-the-art are variants of the Spectral Ewald method described in Ref. \citenum{SpectralEwald_Electrostatics}; a key advantage of our IB approach is the ease of handling other types of boundary conditions that are crucial for modeling confined systems, as we will explore in future work. Some additional numerical results are presented in Appendix \ref{sec:IB-P3M}.

In the quasi-electrostatic approximation assumed here, the electrostatic force is found by solving Poisson's equation for the electrical potential $\phi$,
\begin{equation}
\gradspatial^2 \phi = -\chargedensity/{\epsilon},\label{eq:poisson}
\end{equation}
where the charge density is
\begin{equation}
\varrho(\spatial) = \sum_{i=1}^N \delta(\position_i-\spatial) \charge_i,
\label{eq:delta_funcs}
\end{equation}
with $q_i$ being the charge of ion $i$, and $\epsilon$ the electrical permittivity of the solvent. The resulting electric field is then given by
\begin{align}
\efield %^{\rm P}
=- \bs \nabla_r \phi, \label{eq:eform}
\end{align}
with the electrostatic force on ion $i$ being $\force^{\mathrm{E}}_i = \charge_i \efield(\position_i)$; note that the singular self-induced electric field is defined to be zero.

The P3M approach entails solving for the electrical potential on a grid, and then making local close range corrections to the force in order to treat the ions as point charges. Note that for triply-periodic domains, the Spectral Ewald method described in Ref.~\citenum{SpectralEwald_Electrostatics} is the most accurate technique. However, this method is based on Fourier transforms and is nontrivial to generalize to other boundary conditions and domain configurations. 

In the initial stage of the electrostatic force computation we map the particle charges to a mesh, discretize and solve  Eq.~\eqref{eq:poisson}  on the mesh, and interpolate the resulting field back to the particle locations.  Analogous to the hydrodynamics, the Lagrangian particles interact with the Eulerian electrostatic mesh via interpolation and spreading operators.
\begin{align}
\eparticle^{\mathrm{P}}_i  =&\,
\electrointerp_h(\position_i)\efield  = \int \electrokernel(\position_i - \spatial_h)\efield  
(\spatial_h) d\spatial_h,\label{kernel3}\\
\varrho(\spatial_h) =&\, \electrospread_h(\position_i)\chargevec  = \sum_{i=1}^N \electrokernel(\position_i-\spatial_h) \charge_i,\label{kernel4}
\end{align}
where the integral in Eq.~\eqref{kernel3} is a short-hand notation for a weighted sum over grid points, just as for the hydrodynamic interpolation operator. Here
$\eparticle^{\rm P} = \left\{ \eparticle^{\rm P}_1,\cdots,\eparticle^{\rm P}_i,\cdots,\eparticle^{\rm P}_N\right\}$ is the smoothed value of  $\efield$ 
at the particle locations, $\chargevec=\{\charge_1,\cdots,\charge_i,\cdots,\charge_N\}$ is the vector of particle charges, and $\delta^{\mathrm{es}}$ is the kernel used for the electrostatic grid;
we note that $\delta^{\mathrm{es}}$ can be chosen independently from $\hydrokernel$.
The numerical solution of Eq.~(\ref{eq:poisson}) is discussed in Sec.~\ref{subsec:ElectroSolver}. The smoothed electrostatic force on particle $i$ is then given by 
\begin{align}
\force^{\mathrm{P}}_i  = \charge_i \eparticle_i^{\mathrm{P}} .\label{eqn:poissonforce}
\end{align}

A local correction is needed because a point charge representation of the ions would require an unreasonably fine mesh when solving Eq.~\eqref{eq:poisson}; for finite, non-zero $\Delta r$, the force calculated between two particles, $i$ and $j$, on the mesh  becomes less accurate as the distance between the particles becomes smaller.
This effect becomes particularly severe
when the kernels overlap and the particles no longer appear as points, i.e., $\absposition_{ij}<\psi$, where $\psi$ is on the order of the diameter of the kernel. The P3M method accounts for such situations by replacing the Poisson solution with a direct calculation of the Coulomb force,
\begin{align}
    \force_{ij}^\mathrm{C}= \frac{1}{4\pi \epsilon} \frac{\charge_i \charge_j}{\absposition_{ij}^2}\hat{\position}_{ij},\label{eq:coulomb}
\end{align}
where $\absposition_{ij} = |\position_{ij}|$, and the unit vector $\hat{\position}_{ij} = \position_{ij}/\absposition_{ij}$.
Specifically, the electrostatic force is computed as
\begin{align}
    \force_{i}^\mathrm{E}= \force^\mathrm{P}_i +  \sum_{j\in\Omega_i^\mathrm{E}} \force_{ij}^\mathrm{LC},\label{eq:fullElectro}
\end{align}
where $\Omega_i^\mathrm{E}$ indicates all particles within a range of $\psi$ from particle $i$.
Here the local or near-field correction is
\begin{equation}
\force_{ij}^\mathrm{LC} = \force_{ij}^\mathrm{C} - \force_{ij}^\mathrm{P},\label{eq:LC}
\end{equation}
where $\force_{ij}^\mathrm{P}$ is the force computed by the immersed-boundary method if only the charges $i$ and $j$ were present. As mentioned above, $\psi$ is of the order of the diameter of the electrostatic kernel, and is therefore a function of the cell size. In this work we apply an upper limit of $\psi \leq L/2$, where $L$ is the length of the periodic domain, so that for each pair of charges the local correction is only applied for the nearest periodic image of one of the charges.

Following previous P3M and (Spectral) Ewald methods, the local correction is approximated as a radially-symmetric short-ranged force using
\begin{equation}
\force_{ij}^\mathrm{P} \approx \frac{\charge_i \charge_j}{4\pi \epsilon (\Delta\absspatial)^2} F^\mathrm{P}\left(\frac{\absposition_{ij}}{\Delta\absspatial}\right) \hat{\position}_{ij},
\label{eq:F_LC}
\end{equation}
where the dimensionless scalar function $F^\mathrm{P}(y)$ is tabulated for $0\leq y \leq \psi/(\Delta\absspatial)$ in a pre-computation step, which is done once and only once for a given kernel $\delta^{\mathrm{es}}$. The values used for the results in this work are given in Table~\ref{peskinTable} in Appendix \ref{sec:IB-P3M-Tables}.
This table is computed using two charges in a domain which is large enough that boundary effects are negligible. Because there is a small degree of translational variance, the solution for each separation distance is calculated many times using random placements and orientations of the pair relative to the grid, and the results averaged. We show numerical results for the near-field corrections in Appendix \ref{sec:IB-P3M}.

It is important to emphasize a key difference between the IB approach described here and that used in other P3M-style methods, including the spectral Ewald method \cite{SpectralEwald_Electrostatics}. In the latter, the smearing/smoothing of the delta functions in Eq.~\eqref{eq:delta_funcs} is done at the continuum level, using a smooth kernel function such as a Gaussian. For an unbounded domain, the correction potential $\force_{ij}^\mathrm{LC}$ can easily be computed analytically using Fourier-space integration. The smooth continuum density is then spread to the grid on which the Poisson equation is solved (using, e.g., a spectral method in the Spectral Ewald method or a finite-difference approach as done here), and the solution is interpolated back on the particles just as we do. The spreading and interpolation operations to and from the grid are done to spectral accuracy using the non-uniform FFT in the Spectral Ewald method, while in classical P3M methods it is done using a separate ``window'' or ``charge assignment function,'' such as a power of the sinc function \cite{Hockney:1988:CSU:62815}; a plethora of methods exist with different choices of how this step is done \cite{frenkel2001understanding}. By contrast, in our approach there is only \textit{one} kernel, $\electrokernel$, that is used both to smear the charge and to communicate between particles and grid. In Appendix \ref{sec:IB-P3M}, we study the accuracy of our IB-P3M approach and isolate the contributions to the dominant error coming from the loss of translational and rotational invariance from the Poisson solver and from the spreading and interpolation using a compactly-supported kernel.
\vspace{-4mm}
\subsection{Steric Repulsion}

\begin{figure}[h!]
  \centering
    \includegraphics[width=0.95\textwidth]{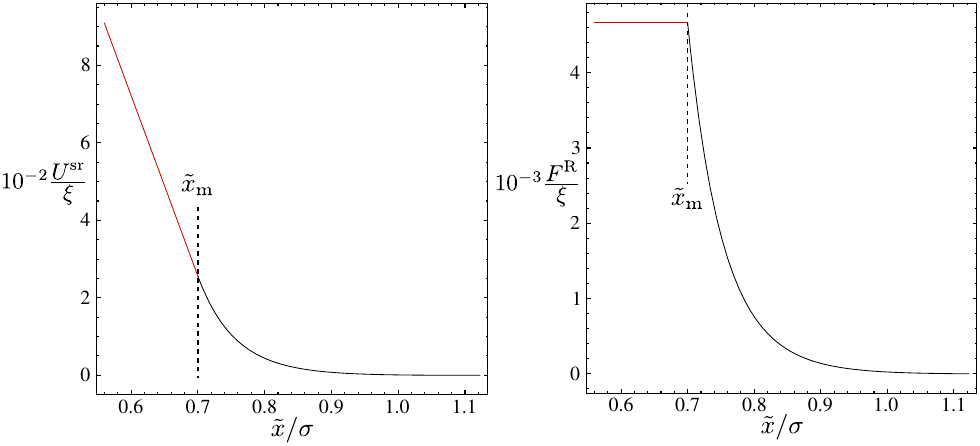}
 \caption{Illustration of short range potential and repulsive force due to Eq.~(\ref{eq:potential}). For ease of viewing, here we have set $\paramposition_m=0.7\sigma$. Note that the actual value used in the simulations, discussed in Sec.~\ref{sec:Results}, is $\paramposition_m=0.25\sigma$.}
\label{fig:potential}
\end{figure}

To compute the short range repulsive force $\force^\mathrm{R}_{ij}$, we model the finite size and excluded volume of the ions using a mollified Weeks-Chandler-Andersen (WCA) interaction potential \cite{Weeks1971}, which is a shifted and truncated Lennard-Jones potential; it has been selected to approximate hard sphere repulsion whilst remaining numerically tractable. This potential is given by
\begin{align}
U^{\rm sr}(\paramposition;\sigma,\xi) =
\begin{cases}
4 \xi \left( \frac{\displaystyle 6 \sigma^6}{\displaystyle  \paramposition_\mathrm{m}^7}- \frac{\displaystyle 12 \sigma^{12}}{\displaystyle  \paramposition_\mathrm{m}^{13}}\right) (\paramposition-\paramposition_\mathrm{m})- 4\xi \left( \frac{\displaystyle  \sigma^6}{\displaystyle  \paramposition_\mathrm{m}^6}- \frac{\displaystyle  \sigma^{12}}{\displaystyle  \paramposition_\mathrm{m}^{12}}\right)+ \xi, \vspace{1mm}& \paramposition \leq \paramposition_{\mathrm{m}}\\
4\xi \left(\left(\frac{\displaystyle\sigma}{\displaystyle\paramposition}\right)^{12} - \left(\frac{\displaystyle\sigma}{\displaystyle\paramposition}\right)^6\right) + \xi, \vspace{1mm}&\paramposition_{\mathrm{m}} < \paramposition < 2^{1/6}\sigma\\
0, & 2^{1/6}\sigma \le \paramposition
\end{cases}\label{eq:potential}
\end{align}
where $\xi$ is the magnitude of the potential, $\sigma$ is the van der Waals diameter, and $\tilde{x}$ is the radial distance between the particles. Note that this potential has been selected to produce a constant force for $\paramposition \leq \paramposition_\mathrm{m}$, equal to that at $\paramposition = \paramposition_\mathrm{m}$, in order to reduce numerical stiffness; specific values of these parameters are discussed in Section~\ref{sec:Results}.
The cutoff at $2^{1/6}\sigma$ ensures that only the repulsive part of the potential is used and shifting the potential by $\xi$ ensures that there is no discontinuity in the potential and in the force at the cutoff. The short range repulsive force is then
\begin{equation}
\force^\mathrm{R}_{ij} = -\hat{\position}_{ij}\,\frac{d}{d \paramposition} U^{\rm sr}_{ij}(\absposition_{ij}; \sigma_{ij},\xi_{ij}),\label{eq:closerange}
\end{equation}
where $\sigma_{ij}$ is the average of the diameters of particles $i$ and $j$,
and in this work we set $\xi_{ij}=\xi$ for all pairs of ions. Note that the repulsion diameter is distinct from the hydrodynamic radius $\radius_t$ discussed in Sec.~\ref{sec:diffusion},  however, both are related to the solvation layer formed around an ion \cite{stokes1948ionic}. The potential and repulsive force are illustrated in Fig.~\ref{fig:potential}.

\section{Details of Numerical Methodology}\label{sec:NumericalMethods}

In this section we discuss the details of the discretization and solution methods of the hydrodynamic and electrostatic\footnote{Note that the hydrodynamic equations are in the quasi-hydrostatic approximation (neglect inertial terms) while the electrostatic equations are in the quasi-electrostatic approximation (neglect magnetic fields).} equations,  Eqs.~(\ref{eq:disc_stokes}) and (\ref{eq:poisson}), and the associated particle time stepping algorithm. These are all implemented in the AMReX
framework described in Ref.~\citenum{zhang2019amrex} and available at \url{https://amrex-codes.github.io/}.

\subsection{Peskin Kernels}\label{sec:peskin}
In sections \ref{sec:fib} and \ref{sec:IonicForces}, we define kernel functions that describe the interaction of the particles with continuous fields; see Eqs.~(\ref{kernel1}), (\ref{kernel2}), (\ref{kernel3}), and (\ref{kernel4}). These functions are represented numerically by Peskin kernels \cite{peskin2002}, which approximate Gaussian functions on a discrete grid. The four-point kernel is defined using
\begin{align}
\delta^\mathrm{Pe}(\zeta_k)= 
\hspace{-1mm}\begin{cases} 
      \frac{\displaystyle 3-2|\zeta_k| + \sqrt{\displaystyle 1+4|\zeta_k| -4|\zeta_k|^{2}} }{\displaystyle 8 \Delta \absspatial}, &  
      \hspace{-2mm} 0 \leq  |\zeta_k|  \leq 1 \vspace{2mm} \\
      \frac{ \displaystyle 5 - 2|\zeta_k| - \sqrt{\displaystyle -7+12|\zeta_k| -4|\zeta_k| ^2}}{{\displaystyle 8 \Delta \absspatial}}, & 
      \hspace{-2mm} 1 <|\zeta_k| \leq 2  \vspace{2mm}\\
       0, &\hspace{-2mm}  2<|\zeta_k| 
   \end{cases},\label{peskin1d}
\end{align}
where $\bs{\zeta} =(\position_i-\spatial)/\Delta \absspatial$, $\zeta_k$ indicates a single Cartesian component of $\bs{\zeta}$, ($\zeta_x$, $\zeta_y$, or $\zeta_z$), and $\Delta r$ is the grid spacing, which is the same in each direction.  
Peskin kernels attempt to maximize certain ideal properties of a continuum Gaussian, notably isotropy and translational invariance; a detailed discussion is given in Ref.~\citenum{peskin2002}. Recently kernels with a greater number of points in their support, including five and six grid points \cite{New6ptKernel}, have been computed. These have improved smoothness and come closer to the ideal case at the cost of increased computation time. Conversely, using a smaller number of points will result in a less accurate kernel but reduced computational cost.

Although it is not necessary to use the same kernels for the hydrodynamic and electrostatic fields, for simplicity in this article we use the same four point kernel in both cases, 
\begin{align}
\hydrokernel(\bs{\zeta})=\electrokernel(\bs{\zeta})=\delta^\mathrm{Pe}(\zeta_x)\delta^\mathrm{Pe}(\zeta_y)\delta^\mathrm{Pe}(\zeta_z).
\end{align}
We further use the same cell size, $\Delta r$ for all fields, although again this is not a necessity.

The discretized interpolation operation for a single Cartesian component of the hydrodynamic velocity ($v_k$) is
\begin{equation}
\hydrointerp_{h,k}(\position_i)\fluidvelabs_k=\Delta\mathcal{V} \sum_{j\in\Omega_i^\mathrm{Pe}}  \hydrokernel(\position_i - \spatial_j)\fluidvelabs_k(\spatial_j),\label{disckernel1}
\end{equation}
where $\Delta \mathcal{V} = (\Delta \absspatial)^3$, and $j\in\Omega_i^\mathrm{Pe}$ indicates summation over discrete grid
points $\spatial_j$ that lie within the support of $\hydrokernel(\position_i - \bs \spatial_j)$ centered on $\position_i$, and on which the discrete velocity is defined. The spreading operation for a single component of force is given by
\begin{equation}
\hydrospread_h(\position_i)\force_k= \sum_{i=1}^N\sum_{j\in\Omega_i^{\mathrm{Pe}}}  \hydrokernel(\position_i - \spatial_j)\absforce_{i,k}.\label{kernel2disc}
\end{equation}
Similarly, the discretized electrostatic interpolation and spreading operators are, respectively,
\begin{equation}
\electrointerp_{h,k}(\position_i)\efieldabs_k= \Delta\mathcal{V} \sum_{j\in\Omega_i^{\mathrm{Pe}}}  \electrokernel(\position_i - \spatial_j)\efieldabs_k(\spatial_j),\label{eq:discfieldinterp}
\end{equation}
and
\begin{equation}
  \electrospread_h(\position_i)\chargevec=\sum_{i=1}^N\sum_{j\in\Omega_i^{\mathrm{Pe}}}  \electrokernel(\position_i - \spatial_j)\charge_{i}.\label{eq:discchargespread}
\end{equation}

As mentioned above, one reason for using Peskin kernels is to maximize translational invariance. However a small degree of variance remains -- the four point kernel used here has a position dependent hydrodynamic radius of $(1.255\pm0.005)\Delta \absspatial$. Were perfect translational invariance achieved, for unbounded/triply periodic domains the stochastic drift term in the particle equation of motion, Eqs.~(\ref{eq:ions}) and (\ref{eq:ions2}), would be zero everywhere, as the divergence of the mobility would be zero \cite{donev_mesoscopic_diffusion:2014}. However the slight lack of invariance of the kernels renders the stochastic drift term necessary even in domains without solid boundaries; this is discussed in detail in Ref.~\citenum{delong2014brownian}.

\subsection{Fluid solver}

\begin{figure}[h!]
 
  \centering
    \includegraphics[width=0.95\textwidth]{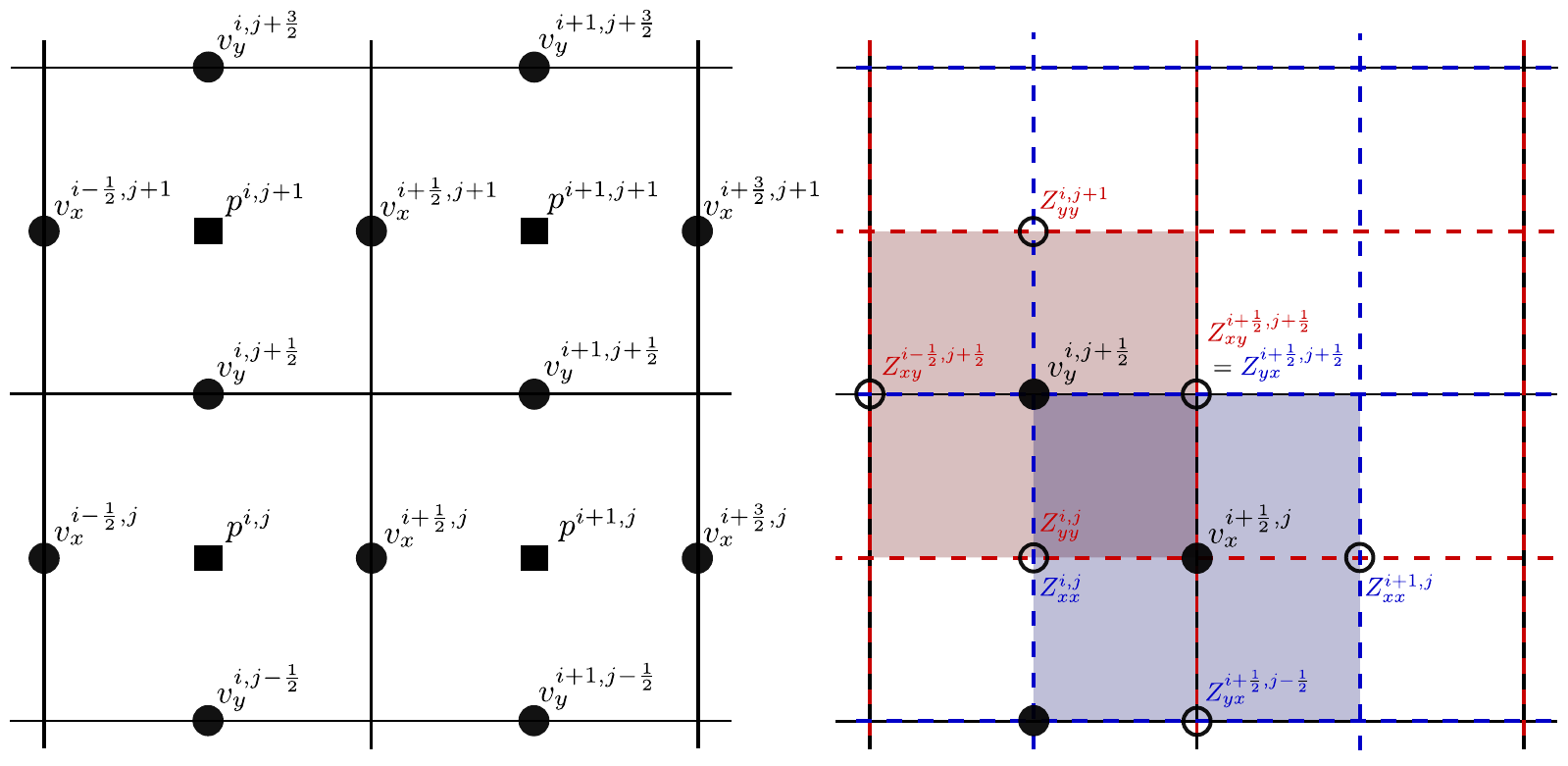}
 \caption{Left: A 2D illustration of the MAC \cite{Cai2014} discretization used for Eq.~(\ref{eqn:discstokes}). Right: 2D illustration of the $\gradstoch\cdot\discretefieldnoise$ term in Eq.~(\ref{eqn:rfd}), which is the discretization of the term $\gradspatial \cdot \fieldnoise$ from Eq.~(\ref{eq:stokes}).
Random numbers, $\discretefieldnoise$, are generated at the faces of control volumes around the points where velocities $\fluidvel$ are defined so that the divergence of the field $\fieldnoise$ can be calculated.}
\label{fig:grid}
\end{figure}

We discretize the fluid velocity on a uniform Cartesian grid with mesh spacing $\Delta \absspatial$.  We use a staggered grid system
with normal velocities and force density defined at cell faces, and pressure defined in cell centers, i.e., a standard marker-and-cell (MAC) discretization \cite{Cai2014}.
Multiple discretizations are used for differential operators representing the gradient, divergence, and Laplacian. Each of these operators use centered second-order differences.
We use $\graddiscrete^{c\rightarrow f}$ to represent the face-centered gradient of a cell-centered field,
$\graddiscrete^{c}$ to represent the cell-centered gradient of a cell-centered field,
$\graddiscrete^{f\rightarrow c}\cdot$ to represent the cell-centered divergence of a face-centered field,
$(\graddiscrete^2)^f$ to represent the face-centered Laplacian of a face-centered field,
and $(\graddiscrete^2)^c$ to represent the cell-centered Laplacian of a cell-centered field.
Finally, $\gradstoch\cdot$ is used to represent the face-centered divergence of a field defined on control volumes corresponding to the shifted (staggered) velocity grid. This is illustrated in two dimensions in Fig.~\ref{fig:grid}. Using these operators, the discrete fluid equation solved at each time step is (see Eqs.~\eqref{eq:disc_stokes} and \eqref{eq:total_f_Stokes}),
\begin{subequations}
\begin{align}
-\eta&(\graddiscrete^2)^f\fluidvel + \graddiscrete^{c\rightarrow f}p = \hydrospread_h\force + \sqrt{\frac{2\eta \kb T}{\Delta t\Delta \mathcal{V}}}\gradstoch\cdot\fulldiscretenoise  + 
\left[
\hydrospread_h\left(\position+\frac{\Delta_{\rm R}}{2}\rfdnoise \right) - \hydrospread_h\left(\position-\frac{\Delta_{\rm R}}{2}\rfdnoise  \right)
\right]\frac{\kb T}{\Delta_{\rm R}}\rfdnoise ,\label{eqn:rfd}\\
&\nonumber\\
&\hspace{70mm}\graddiscrete^{f\rightarrow c}\cdot\fluidvel = 0.
\end{align}\label{eqn:discstokes}%
\end{subequations}
Here,
$\fulldiscretenoise$ represents
Gaussian random numbers of mean zero and variance one, 
defined such that their divergence can be calculated at the locations where $\fluidvel$ is defined. This is illustrated in two dimensions in Fig.~\ref{fig:grid}, where $\fulldiscretenoise$ is stored on cell nodes and centers. In three dimensions $\fulldiscretenoise$ is stored on cell edges and centers.
This is described in detail in Ref.~\citenum{nonaka2015low}. The vector 
$\rfdnoise  = \left\{\rfdnoise_1,\cdots,\rfdnoise_i,\cdots,\rfdnoise_N  \right\}$ consists of independent random vectors \mbox{$\rfdnoise_i = \left(\rfdnoiseabs_{i,x},\rfdnoiseabs_{i,y},\rfdnoiseabs_{i,z}\right)$}, where each $\rfdnoiseabs$ is an independent Gaussian random number of mean zero and variance one.

The second line of Eq.~(\ref{eqn:rfd}) represents the thermal forcing term given by Eq.~(\ref{eqn:thermforce}). As discussed in Ref.~\citenum{delong2014brownian}, it is essentially a finite difference representation of the divergence of the spreading operator. It is implemented by spreading a random force, $\kb T\rfdnoise/\Delta_{\rm R}$, at locations randomly offset from the particle positions, $\position+\Delta_{\rm R}\rfdnoise/2$, and the negative of that force at positions $\position-\Delta_{\rm R}\rfdnoise/2$. The distance $\Delta_{\rm R}$ should be smaller than the length scale over which $\hydrospread$ varies, but large enough to avoid issues due to numerical roundoff. In this article we have used the value $\Delta_{\rm R} = 10^{-4} \Delta \absspatial$. 

The system of discrete equations formed by Eq.~(\ref{eqn:discstokes}) can be solved by a range of approaches, here we have used a preconditioned  Generalized Minimal RESidual (GMRES) method \cite{Cai2014,saad1986gmres}.

\subsection{Electrostatic solver}\label{subsec:ElectroSolver}

The electrostatic potential and charge density, $\phi$ and $\varrho$, are defined on a cell centered grid. The Poisson equation, Eq.~(\ref{eq:poisson}), is discretized by
\begin{equation}
(\graddiscrete^2)^c \phi = -{\chargedensity}/{\epsilon}.\label{eq:poissondisc}
\end{equation}
In this case, the resulting system of equations are solved using the geometric multigrid \cite{press2007numerical} method.  

The resulting electric field is also stored at cell centers, and is found by
\begin{equation}
 \efield =- \graddiscrete^{c} \phi.\label{eq:fielddisc}
\end{equation}
The choice of a cell centered differencing in Eq.~\eqref{eq:fielddisc} has some advantages over other options that are worth noting. Importantly, it can be shown\footnote{We thank Charles Peskin for sharing with us an analytical proof of this property.}
that Eq.~\eqref{eq:fielddisc}, together with the IB interpolation and spreading, ensures that, at least for periodic domains, Newton's third law is obeyed. That is, the force on ion $i$ due to ion $j$ is equal and opposite to that on ion $j$ due to ion $i$,\footnote{Note that this force is, however, not strictly central. That is, the Weak Law of Action and Reaction (Newton's Third Law) is obeyed but not the Strong Law of Action and Reaction, which requires that the forces act along the line joining the particles.} and momentum is conserved.\footnote{Note that in the overdamped limit there is no momentum conservation per se, however, the steady Stokes equation would not be solvable in triply-periodic domains if the forces did not sum to zero.} This implies that there is strictly no self-force of one ion on itself, which is an important property desirable in any P3M method.
It may appear more natural for staggered grids to compute electric field not at cell centers using $\graddiscrete^{c}$, but on cell faces using $\graddiscrete^{c\rightarrow f}$, in particular because
$(\graddiscrete^2)^c = \graddiscrete^{f\rightarrow c}\cdot \graddiscrete^{c\rightarrow f}$. However, this choice does not lead to equal and opposite action/reaction forces and can generate spurious self-forces.

One thing that is lost with the discretisation employed in Eq.~\eqref{eq:fielddisc} is that the force is no longer a negative gradient of an electrostatic potential. That is, the IB-P3M approach followed here cannot be used for applications where both forces and energies matter and need to be consistent with each other; in BD or \SHIBA we only need forces, and  since the system is isothermal there are no issues regarding energy conservation. It is possible to ensure consistency between electrostatic forces and energy if one uses the derivative of the Peskin kernel to interpolate $-\phi$ at the particle positions to obtain the electric field/force. However, this requires using a smoothly-differentiable kernel such as the 6pt Peskin kernel \cite{New6ptKernel}, and also does not lead to momentum conservation.

When applying the close range correction to the electrostatic solution (and the steric repulsion), near linear scaling is maintained by using the neighbor list feature included with AMReX.
\subsection{Temporal Algorithm}\label{subsec:TemporalAlgorithm}
The time stepping scheme employed in \SHIBA builds on that of the FIB method with the addition of electrostatic forces and dry diffusion. 
A time step is then defined by the following four steps:
\begin{enumerate}
\item The charge density $\varrho$ is computed on the grid using the spreading operation defined by Eqs.~(\ref{kernel4}) and (\ref{eq:discchargespread}). Equation~(\ref{eq:poissondisc}) is then solved using the geometric multigrid method, and Eq.~(\ref{eq:fielddisc}) used to obtain the coarse electric field, $\efield$.
\item This electric field is interpolated to the particle locations using the operation defined by Eqs.~(\ref{kernel3}) and (\ref{eq:discfieldinterp}). The corresponding force on the particles is found using Eq.~(\ref{eqn:poissonforce}). For particles at close range, this force is corrected as per Eqs.~(\ref{eq:coulomb}) and (\ref{eq:fullElectro}). Close range WCA interactions are also calculated in this step, and the total force on the particle calculated as per Eqs.~(\ref{eq:totalforce})-(\ref{eq:closerange}), including the effect of any external fields.
\item The force on the particles, $\force$, is spread to the grid storing the force density using the operations defined in Eqs.~(\ref{kernel2}) and (\ref{kernel2disc}), recovering $\forcedensity^\mathrm{s}$. The random finite difference term is spread as per the second line of Eq.~(\ref{eqn:rfd}), yielding the thermal forcing $\forcedensity^\mathrm{th}$. The resulting system, Eq.~(\ref{eqn:discstokes}), is solved via the GMRES method to compute the velocity at time step $n$, $\fluidvel^n$. 
\item The fluid velocity $\fluidvel^n$ is interpolated to particle locations using the operations defined in Eqs.~(\ref{kernel1}) and (\ref{disckernel1}) to obtain the ``wet'' component of the particle velocities, Eq.~\eqref{eq:ions}. The temporal discretization of the particle diffusion is then given by the midpoint update scheme
\begin{eqnarray}
\position_i^{n+1/2,\star} &=& \position_i^n + \frac{\Delta t}{2}(\hydrointerp_h(\position_i^n)\fluidvel^n), \label{step1}\\
\position_i^{n+1} &=& \position_i^n+\Delta t \left[\hydrointerp_h(\position_i^{n+1/2,\star})\fluidvel^n + \frac{D_i^{\rm dry}}{\kb T}\force_i^n \right.
+\left. \sqrt{\frac{2D_i^{\rm dry}}{\Delta t}} \bs{W}_i^n\right].\label{step2}
\end{eqnarray}
As discussed in Sec.~\ref{sec:fib}, the purpose of the midpoint update is to incorporate the part of the stochastic drift not accounted for by the random finite difference force density $\forcedensity^\mathrm{th}$.
\end{enumerate}

This algorithm is first order in time, as illustrated in Ref.~\citenum{delong2014brownian}, where higher order schemes are also discussed.

\section{Numerical Results}\label{sec:Results}

In this section we test the \SHIBA algorithm by comparison with theoretical results for the radial pair correlation function and electrical conductivity. Additionally, we analyze the effect of changing the ratio of wet and dry diffusion.

In our numerical tests we model a 1:1 strong electrolyte solution, similar to salt water, with species labeled $A$ and $B$, with charges $\charge = \charge_A = - \charge_B = 1.6\times 10^{-19}~ \mathrm{C}$. 
The solvent is taken to be water at $T=295$~K, viscosity $\eta = 0.01~ \mathrm{g/(cm\, s)}$, and permittivity $\epsilon = \epsilon_r\epsilon_0$, with relative permittivity $\epsilon_r=78.3$ where $\epsilon_0$ is the vacuum permittivity. The diffusion coefficients of the ions in water are $D_A^{\rm tot} = 1.17\times 10^{-5}~ \mathrm{cm^2/s}$ and $D_B^{\rm tot} = 1.33\times 10^{-5}~ \mathrm{cm^2/s}$, corresponding to hydrodynamic radii of $a_t=0.185~\mathrm{nm}$, and $a_t=0.162~\mathrm{nm}$, respectively. In all cases the close range potential parameters are $\xi = 10^{-16}$ ergs, $\sigma = 0.4~\mathrm{nm}$, and $\paramposition_\mathrm{m} = 0.1~\mathrm{nm}$.
Note that the above parameters yield Schmidt numbers of 854 and 752 for species $A$ and $B$, respectively. This justifies the use of the infinite Schmidt number approximation discussed above.

For each problem the time step, $\Delta t$, 
was selected by successive refinement until a negligible change in the result was observed. Note that the ion diffusive time scale, $a_t^2/D$, is on the order of 10 picoseconds. The time steps used for the simulations below were typically constrained to the order of 0.1~ps by the stiffness of the steric and electrostatic interactions. We are currently examining how the time step may be increased by varying the parameters of these potentials.

In each case the system size was selected to avoid significant finite size effects. When considering electrostatic interactions, the relevant length scale is the Debye length \cite{debye1923theory,robinson2012electrolyte},
\begin{equation}
\lambda_D = \sqrt{\frac{\epsilon \kb T}{\sum_{j=1}^{N_s} n_j \charge_j^2}},\label{eq:Debye}
\end{equation}
where $N_s$ it the number of species, and $n_j$ is the number density of species $j$. This is a measure of how far electrostatic effects persist before they are screened by clouds of opposite charges.
For hydrodynamic effects, it should be noted that the diffusion coefficient of an isolated particle in a triply-periodic domain has well-known strong finite-size corrections of order $a_w/L$ \cite{ISIBM}, where $L$ is the length of the cubic domain. This is because the mobility of an isolated particle is determined by applying a force on it, and a force monopole interacts with its periodic images with $1/r$ hydrodynamic interactions. However each ion diffuses with its Debye cloud, which experiences an equal but opposite force, so that an ion interacts with its periodic images hydrodynamically as a force dipole \cite{ScreenedHydro_Electrolytes}, which has a finite size correction of order $(a_w/L)^3$. We are not aware of any systematic analysis of finite-size effects in the literature; here we have systematically increased the system size until no significant change in the result is observed.

\subsection{Radial Pair Correlation Function}\label{subsec:GofR}

We first validate the equilibrium properties of the \SHIBA method by measuring radial pair correlation functions between ions of like and opposite charge.  In general, the radial pair correlation function is the normalized time-average density of particles as a function of radius from an arbitrary reference particle.  For a binary system, the pair correlation function between species $\alpha$ and $\beta$ is defined as \cite{frenkel2001understanding}
\begin{align}
    &g_{\alpha\beta}(\radvar) =\lim_{\tau\to\infty}\frac{\mathcal{V}}{N_{\alpha\beta}(N_{\alpha\beta}-1) 4\pi\radvar^2 \tau} 
    \int_0^\tau  \sum_{i,j,i\neq j}^N \delta(\radvar - \absposition_{ij}) 
    ~\delta_{\alpha,s_i}\delta_{\beta,s_j}  dt,\label{eq:gofr}
\end{align}
where $\mathcal{V}$ is the system volume, $s_i$ and $s_j$ are the species of particles $i$ and $j$, $N_{\alpha\beta}$ is the number of species pairs, and $\absposition_{ij}$ is the radial distance between particles $i$ and $j$.
In practice, for each snapshot in time, for each ion of type $\alpha$, we count the number of ions of type $\beta$ in thin spherical shells with a specified bin width, and compute a number density in each bin by dividing by the volume of the shell.
We average the results over all $\alpha$ ions and over many time steps, and normalize the result with the average number density of $\beta$.

For low to moderate ion concentrations Debye-H\"uckel theory \cite{debye1923theory,robinson2012electrolyte} gives
\begin{align}
    g_{\alpha\beta}(\radvar) \approx \exp( - U_{\alpha\beta}(\radvar)/\kb T ),\label{eqn:grapprox}
\end{align}
where the potential is
\begin{align}\label{eq:PairCorrPotential}
    U_{\alpha\beta}(\radvar) =  \frac{\charge_\alpha \charge_\beta}{4\pi\epsilon} \frac{e^{- \radvar/\lambda_D}}{\radvar} + U_{\alpha\beta}^\mathrm{sr}(\radvar).
\end{align}
The first term in Eq.~(\ref{eq:PairCorrPotential}) is the screened Coulomb potential \cite{debye1923theory,robinson2012electrolyte} and the second is the short-ranged repulsion.

\begin{figure}[h!]
 
  \centering
    \includegraphics[width=0.95\textwidth]{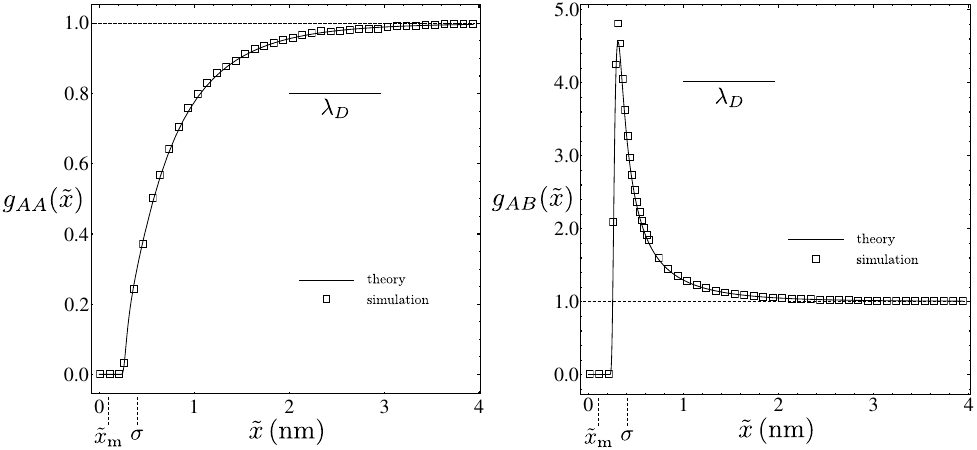}
 \caption{Radial distribution for a molarity of 0.1~M. The right plot shows the pair correlation for ions of like charge; also indicated is the Debye length. The left plot shows the pair correlation for ions of opposite charge. The solid line shows the approximate analytical solution given by Eq.~(\ref{eqn:grapprox}), and the squares show the numerical result from DISCOS. These results were collected using a bin width of 0.025 nm, with a sample size of 50,000 -- the simulation was run for 100,000 steps, with the first 50,000 steps used for equilibration. This sample size resulted in negligible statistical error. For visual clarity, only every fourth point is displayed. The grid size was set such that the total diffusion of species $A$ was 14.7\% wet and 85.3\% dry, and for species $B$, 12.9\% wet and 87.1\% dry. This corresponds to $\Delta r = 1\textrm{nm}$, with a $32\times32\times32$ cell periodic domain, and 3946 ions. The time step was $\Delta t =0.1$~ps. \vspace{13mm}}

\label{fig:gr1}
\end{figure}

\begin{figure}[h!]
 
  \centering
    \includegraphics[width=0.95\textwidth]{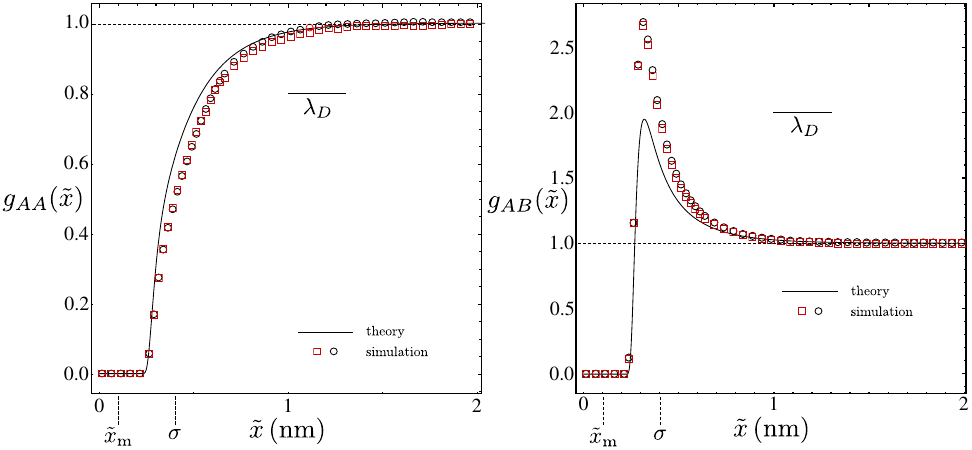}
\caption{
Radial distribution for a molarity of 1.0M. The right plot shows the pair correlation for ions of like charge; also indicated is the Debye length. The left plot shows the pair correlation for ions of opposite charge. The solid line shows the approximate analytical solution given by Eq.~(\ref{eqn:grapprox}), and the boxes and circles show numerical result from DISCOS. The numerical parameters are the same as those described for Fig.~\ref{fig:gr1}, except that two different grid resolutions are shown. The black circles show the case where the grid size was set such that the total diffusion of species $A$ was 14.7\% wet and 85.3\% dry, and for species $B$, 12.9\% wet and 87.1\% dry. This corresponds to $\Delta r = 1\textrm{nm}$, with a $16 \times 16\times 16$ cell periodic domain, and 4932 ions. The red squares show the case where the grid size was set such that the total diffusion of species $A$ was 58.9\% wet and 41.1\% dry, and for species $B$, 51.8\% wet and 49.2\% dry. This corresponds to $\Delta r = 0.25\textrm{nm}$, with a $32 \times 32\times 32$ cell periodic domain, and 618 ions.\vspace{6mm}
}
\label{fig:gr2}
\end{figure}

In Figs.~\ref{fig:gr1} and \ref{fig:gr2}, we compare the approximate theoretical expression of Eq.~(\ref{eqn:grapprox}) with results obtained from \SHIBAe. In Fig.~\ref{fig:gr1}, the comparison is shown for molarity (moles of cation or anion per liter of solvent) of 0.1~M; excellent agreement is observed.
Fig.~\ref{fig:gr2} shows the comparison for a molarity of 1.0 M for two different ratios of wet and dry diffusion. A negligible difference is observed between second and third simulations; this is unsurprising as we can see from Eq.~(\ref{eqn:grapprox}) that the pair correlation function does not depend on the hydrodynamic properties of the solvent. Reasonable agreement is observed, with the largest deviation observed in the peak of the opposite charge result. We note that Eq.~(\ref{eqn:grapprox}) is derived in the low concentration limit so we expect decreasing agreement with increasing concentration. At the higher concentrations the Percus-Yevick and hypernetted chain (HNC) approximations are more accurate~\cite{hansen1990theory}.

\subsection{Electrical Conductivity}\label{subsec:Conductivity}

For an applied electric field of magnitude $E$ the total current density in an electrolyte solution is
\begin{equation}
\mathcal{I} = \sum_{j=1}^{N_s} n_j \charge_j^2 \mu_j E,
\end{equation}
where $\mu_j$ is the mobility of species $j$. The conductivity is defined by Ohm's law as $\mathcal{C} = \mathcal{I}/E$ and
from the Nernst-Einstein relation \cite{robinson2012electrolyte} the mobility in the limit of infinite dilution is
$\mu_j^0 = D_j^\mathrm{tot}/\kb T$. 
In this limit, the uncorrected total conductivity $\mathcal{C}^0$ is the sum of the species conductivities $\mathcal{C}_i^0$,
\begin{equation}
    \mathcal{C}^0 = \sum_{j=1}^{N_s} \mathcal{C}_j^0
    = \sum_{j=1}^{N_s} \frac{n_j \charge_j^2 D^\mathrm{tot}_j}{\kb T}.
\end{equation}
Debye-H{\"u}ckel-Onsager (DHO) theory gives the electrolyte conductivity \cite{OnsagerFuoss1932,robinson2012electrolyte} as the sum of the uncorrected total conductivity, electrophoretic contribution, and a contribution due to the relaxation effect,
\begin{equation}\label{eq:FullDHO}
\mathcal{C} = \mathcal{C}^0
+ \mathcal{C}^{\rm ep} + \mathcal{C}^{\rm relx}.
\end{equation}
See Ref. \citenum{donev2019fluctuating} for the relationship of this theory to fluctuating hydrodynamics.

The electrophoretic contribution is due to screening charges of opposite sign about each ion imparting a retarding viscous stress on the ion, given by
\begin{equation}
\mathcal{C}^{\rm ep} =
-\sum_{j=1}^{N_s} \frac{n_j \charge_j^2}{6\pi \eta \rho \lambda_D}
~\frac{\lambda_D}{\lambda_D + a_j^\mathrm{DH}}.
\end{equation}
The last term is a Debye-H{\"u}ckel correction to account for the finite ion size, which we take to be $a_j^\mathrm{DH} = 0.4$~nm for both ions.
The relaxation effect refers to the average force experienced by an ion from its asymmetric ionic cloud relaxing due to thermal fluctuations.
For binary electrolytes the contribution due to the relaxation effect is
\begin{equation}
\mathcal{C}^{\rm relx} =
-\sum_{j=1}^2
\frac{\mathcal{C}^0}{12 \pi (2 + \sqrt{2})} 
\frac{\charge_j^2}{\epsilon \kb T \lambda_D}
~\frac{\lambda_D}{\lambda_D + a_j^\mathrm{DH}}.
\end{equation}
For $a_j^\mathrm{DH} \ll \lambda_D$ both the electrophoretic and relaxation contributions go as the square root of the ionic concentration.

The $\mathcal{C}^0$ term contains no hydrodynamic interactions between ions; it is therefore correctly captured by a simulation with any wet/dry diffusion ratio. The same is true for the $\mathcal{C}^{\rm relx}$ term, as this arises entirely from electrostatic interactions which are well-resolved using the P3M approach. The $\mathcal{C}^{\rm ep}$ term arises from the collective hydrodynamic effect of an ion and its screening cloud of opposite charge, it is therefore only completely captured by a simulation using 100\% wet diffusion; the relative importance of this effect is discussed in the next section.

\begin{figure}[h!]
 
   \centering
    \includegraphics[width=0.47\textwidth]{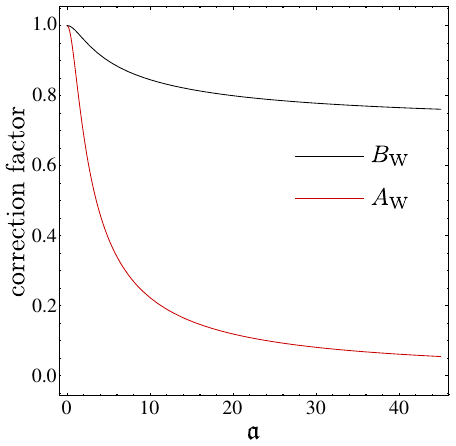}
 \caption{Wien corrections to the DHO relaxation effect and electrophoretic contribution as a function of $\aWien = (\charge \lambda_D E)/(\kb T)$.}
\label{fig:DHO_corrections}
\end{figure}

When the applied electric field is sufficiently strong the distortion of the ionic clouds results in an increased electrical conductivity, which is known as the first\footnote{Note that there is a ``second Wien effect'' which applies to weak electrolytes.} Wien effect. For a 1:1 strong electrolyte solution the Wien corrections to the conductivity are~\cite{harned1958physical}
\begin{align}
    \mathcal{C}_\mathrm{W}^\mathrm{relx}(\aWien) &= A_\mathrm{W}(\aWien)~\mathcal{C}^\mathrm{relx},
    \\
    \mathcal{C}_\mathrm{W}^\mathrm{ep}(\aWien) &= B_\mathrm{W}(\aWien)~\mathcal{C}^\mathrm{ep},
\end{align}
where
\begin{align}
    A_\mathrm{W}(\aWien) = \frac{3(1+\sqrt{1/2})}{2 \aWien^3}
    \Big( \aWien \sqrt{1+\aWien^2} - \sqrt{2}\aWien 
    + \arctan(\sqrt{2}\aWien) - \arctan(\aWien/\sqrt{1+\aWien^2}) \Big),
\end{align}
\begin{align}
    B_\mathrm{W}&(\aWien) = \frac{1}{\sqrt{2}} + \frac{3}{8\aWien^3}
    \Big( 2 \aWien^2 \mathrm{arcsinh}(\aWien) - \aWien \sqrt{1+\aWien^2} + \sqrt{2}\aWien 
    &-(1+2\aWien^2)\left[\arctan(\sqrt{2}\aWien) - \arctan(\aWien/\sqrt{1+\aWien^2})\right]
    \Big),
\end{align}
and $\aWien = (\charge \lambda_D E)/(\kb T)$ is a dimensionless parameter for the electric field strength.
These correction factors are shown in Fig.~\ref{fig:DHO_corrections}.
Note that $A_\mathrm{W}(\aWien) \rightarrow 0$ and $B_\mathrm{W}(\aWien) \rightarrow 1/\sqrt{2}$ as $\aWien \rightarrow \infty$.

In the presence of an applied electric field, the conductivity vector can be computed by
\begin{equation}
\bs{\mathcal{C}} = \lim_{\tau\to\infty}\frac{1}{E \mathcal{V} \tau} \left[\bs{\mathfrak{Z}}(\tau) - \bs{\mathfrak{Z}}(0)\right],\label{eq:simpleCond}
\end{equation}
where the ion polarization (displacement of the ``center of charge'') is
\begin{align}
\bs{\mathfrak{Z}}(t) = \sum_{i=1}^N \charge_i \position_i(t).
\end{align}

An alternative way to measure conductivity in the absence of an applied field is given by the Einstein-Helfand formula \cite{jardat1999transport}, in terms of the long-time diffusion coefficient of the center of charge,
\begin{align}
\bs{\mathcal{C}}^{\mathrm{EH}} = \lim_{\tau\to\infty} \frac{1}{6 \kb T \mathcal{V}\tau} \int_0^\tau \left[\bs{\mathfrak{Z}}(t) - \bs{\mathfrak{Z}}(0)\right]^2 dt.\label{eq:EHCond}
\end{align}
This is derived from the Green-Kubo relation \cite{kubo1966fluctuation,jones2012adaptive}, and is discussed in Refs.~\citenum{jardat1999transport} and \citenum{nieszporek2016calculations}. 

Formally, Eqs.~(\ref{eq:simpleCond}) and (\ref{eq:EHCond}) hold when $\tau$ is large enough to ensure that the short time correlations have decayed; increasing $\tau$ beyond this point has no effect on the statistical convergence, as the variance of $\bs{\mathfrak{Z}}$ does not decrease with time. An accurate result can be obtained by increasing $N$, or averaging an ensemble of simulations. We find that the mean-square displacement of the center of charge is perfectly linear in time on our overdamped time scales for all $\tau$ examined, so the conductivity is not sensitive to the choice of $\tau$. The correct result can therefore be obtained by setting $\tau \sim \Delta t$ to maximize the statistical sample size. To ensure de-correlation between simulation measurements the conductivity was sampled at intervals of 10 ps, which is on the order of the ion diffusion time, when calculating the time average in Eq.~(\ref{eq:simpleCond}). For Eq.~(\ref{eq:EHCond}), the displacement in $\bs{\mathfrak{Z}}$ was measured and then reset every 10 ps. Note that this has no impact on the result itself, but it influences the calculation of the statistical error.

Three sets of conductivity simulation measurements were performed at a concentration of 0.1~M: (i) Measurements with zero applied field using Eq.~(\ref{eq:EHCond}), (ii) using a weak applied field of $10^5~\mathrm{V/cm}$ ($\mathfrak{a}=0.377$), and (iii) using a strong applied field with a value of $10^7~\mathrm{V/cm}$ ($\mathfrak{a}=37.7$), each using Eq.~(\ref{eq:simpleCond}). In each case a range of diffusion ratios, i.e.~values of $D^\mathrm{wet}$ and $D^\mathrm{dry}$, were tested. Tests using case (i) with Eq.~(\ref{eq:EHCond}) were also performed at concentrations of 0.0115~M, 0.5~M, and  2~M. The results are summarized in Table~\ref{table:conductivity_data} and illustrated in Fig.~\ref{fig:cond}.

 All simulations were performed in a cubic domain with
 length $L = 10.043$~nm. For concentrations 0.0115~M, 0.1~M, 0.5~M, and 2~M, the domain contained 14, 122, 610, and 2440 ions, respectively. For comparison, a molecular dynamics simulation of this system would contain roughly 30,000 water molecules.
 
To vary the wetness, the computational domain was divided into $8^3, 16^3, 32^3,$ or $64^3$ grid cells for both the Stokes and Poisson solvers; simulations with fully dry diffusion used a $32^3$ grid for the Poisson solver.  The $8^3$ grid cell cases correspond to wetness percentages of $11.8\%$ and $10.3\%$ for ion A and B, respectively, while the $64^3$ grid cell cases correspond to wetness percentages of $93.8\%$ and $82.6\%$. Most simulations for molar concentrations of 0.5~M and below used a time step size of 0.1~ps, while for the high molarity 2~M cases $\Delta t$ was typically between 0.01 and 0.02~ps. It is important to note that while a Stokes grid as coarse as $8^3$ does little to accelerate the far-field calculations for nearest image particles, the use of a periodic Stokes solver is crucial to account for the hydrodynamic interactions with periodic image particles; the same applies also to the Poisson solver.

In all cases, good agreement is observed between the results with the highest $D^\mathrm{wet}$ value and the theoretical prediction, for both high and low field cases, and at all concentrations.
The weak and strong field results are in particularly good agreement with the theory for both fully dry and fully wet simulations. Note that DHO theory may not be accurate at high molarities, especially since we do not tune the value of the finite-size radius $a^\mathrm{DH}$ for either species.
The dry ($D^\mathrm{wet}$=0) cases also compare reasonably well with the theoretical result with the electrophoretic component removed, in agreement with the discussion above.

For the weak field simulations, the strength of the field has been set such that the Wien effect is negligible; we therefore expect the same results for the weak and zero field cases. Interestingly, a difference of about 3\% is observed for the highest wet percentage, and 5\% for the lowest. The cause of this discrepancy is not apparent, but warrants further investigation.

\begin{table*}[htb]
\begin{center}
\begin{tabular}{|l|c|c|c|c|c|c|c|}
\hline
\multicolumn{1}{|c}{}  & \multicolumn{1}{c}{} & \multicolumn{5}{c}{\makecell{Measured in case wet\% $A$, wet\% $B$}}& \\
\hline
   &Theory& 0\% & 11.75\%\ & 23.5\% & 46.9\% & 93.8\%&Theory\\
  &No EP  & 0\% & 10.33\% & 20.65\% & 41.3\% & 82.6\%&With EP\\
  \hline
    \multicolumn{1}{|c}{} & \multicolumn{7}{c|}{Zero $E$-field} \\
\hline
  0.01147~M & 0.106& 0.1086(18)&  0.1013(6) & 0.1020(21)  & 0.1002(21) & 0.1010(18) &0.100\\
 %& & 0.5& 0.25 &  1& 1& 1 & 1 &\\
  \hline
  0.1~M & 0.897&   0.946(12) &   0.869(27)&  0.837(21) & 0.813(9) & 0.803(12)&0.776\\
%           & &   0.5 &  0.25 &   1 &  1& 1 & 1&\\
    \hline
0.5~M &4.33& 4.70(6)& 4.21(9)&3.95(6)&  3.63(6) &  3.54(15)&3.34  \\
 %&&1&0.01 & 0.05&1&  1 &  1&  \\
    \hline
2~M & 16.8& 18.5(6) & 16.7(9)& 14.8(9) &12.8(6)&  11.6(9) &  11.4  \\
%& & 0.5 & 0.025 &0.2& 0.2 &0.2&  0.1 &  \\
 \hline
 \multicolumn{1}{|c}{} & \multicolumn{7}{c|}{Weak $E$-field} \\
\hline
 0.1~M & 0.897 & 0.898(6)&  0.831(6) & 0.785(6) & 0.785(9) & 0.777(6) & 0.776\\
 %&& 0.5 &  &  &  &1&1& \\
  \hline
 \multicolumn{1}{|c}{} & \multicolumn{7}{c|}{Strong $E$-field} \\
\hline
0.1~M &0.943& 0.943(3) &  0.870(3)& 0.851(3) & 0.841(3) & 0.837(3)&0.850 \\
 %&& 0.5 &  &  &  &1&1& \\
  \hline
\end{tabular}
\end{center}
\caption{\label{table:conductivity_data} Conductivity simulation measurements in siemens per meter for zero $E$-field (by Einstein-Helfand, Eq.~(\ref{eq:EHCond})) and weak and strong $E$-field (center of charge displacement, Eq.~(\ref{eq:simpleCond})). 
The uncertainty notation is defined such that the numbers in parenthesis indicates the error bar for the final digits, e.g., 
$0.898(6)$ indicates $0.898\pm0.006$, and $0.946(12)$ indicates $0.946\pm0.012$. The right column is calculated from DHO theory including Wien effect corrections for the strong $E$-field cases. For comparison with the 0\% wet case, the theory prediction without the electrophoretic (EP) correction is also included in the table. A plot of these results is given in Fig.~\ref{fig:cond}.
}\label{condTable}
\end{table*}

\subsection{Effect of wet vs. dry diffusion}\label{subsec:wetdry}

\begin{figure}[h!]
   \centering
    \includegraphics[width=0.475\textwidth]{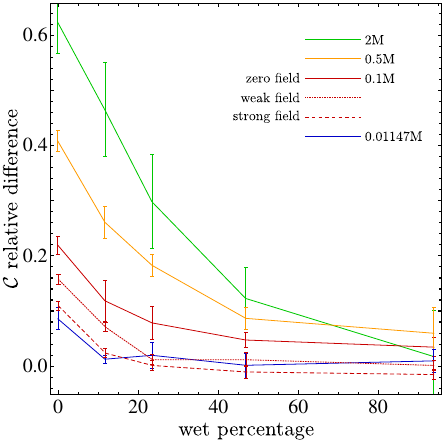}
 \caption{Relative difference in conductivity from DHO theory including Wien corrections for the strong field case), as a function of wetness percentage, for a range of molarities; based on data from Table~\ref{condTable}.}
\label{fig:cond}
\end{figure}

The results in the previous section show that while some wet diffusion is needed to capture the electrophoretic contribution to conductivity, in most cases increasing the wet diffusion above 50\% has a relatively small impact. Furthermore the relative effect depends on concentration, with wet diffusion making a greater contribution at higher molar concentrations; this can be clearly seen in Fig.~\ref{fig:cond}. As mentioned in Sec.~\ref{sec:diffusion},  reducing the hydrodynamic resolution (by reducing $D^\mathrm{wet}$) neglects short range hydrodynamic interactions. It seems clear that short range hydrodynamic interactions should increase in importance as concentration increases, i.e., as the particles are on average closer together. However it must be noted that this will also increase the relative importance of electrostatic interactions, which decay more rapidly than hydrodynamic influence.

\begin{figure}[h!]
   \centering
    \includegraphics[width=0.95\textwidth]{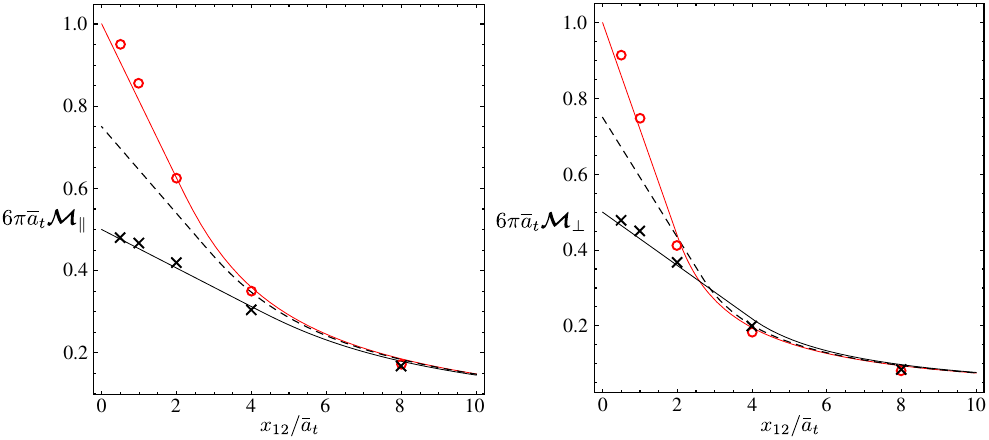}
 \caption{{Left: Mobility of particle 1 when a force is applied on particle 2, in the direction parallel to the vector connecting particles 1 and 2. Right:  Mobility of particle 1 when a force is applied on particle 2, in the direction perpendicular to the vector connecting particles 1 and 2. The red circles are values measured from a 100\% wet simulation, the black crosses from a 50\% wet simulation. In each case the values are normalized by the particle self mobility, and the separation is shown in terms of the average particle radius of species $A$ and $B$, i.e., $\bar{a}_t=(a_t^A+a_t^B)/2$, $\bar{a}_w=(a_w^A+a_w^B)/2$, $\bar{a}_d=(a_d^A+a_d^B)/2$. Each set of data is compared to the result from the RPY tensor calculated using $a_w$, shown with the red and black solid lines. Also shown for comparison is the RPY result corresponding to the 75\% wet case (dashed line). The approach of using the average particle radius has been used for simplicity because the two species are of similar size; we note that a polydisperse version of the RPY tensor is described in Ref.~\citenum{zuk2014rotne}. Simple fits of the pair mobility for the 6-point Peskin kernel is given in Ref.~\citenum{balboa2017hydrodynamics}.}
}
\label{fig:mobcompare}
\end{figure}

To estimate the importance of these effects, and obtain an estimate for the required $D^\mathrm{wet}$ value for a given concentration, we consider two ions, 1 and 2. In Fig.~\ref{fig:mobcompare}, the effect of dry diffusion on the particle mobility matrix is illustrated as a function of particle separation. The mobility parallel and perpendicular to the vector connecting the particles is shown, measured from 100\% wet and 50\% wet simulations. These are compared to the solution obtained from the RPY tensor, Eq.~(\ref{eq:rpy}). As discussed above, increasing the percentage of dry diffusion amounts to neglecting short range hydrodynamic interactions; as per Eqs.~(\ref{eq:diffTotal})--(\ref{diffSum}), changing from 100\% wet to 50\% wet is equivalent to doubling the hydrodynamic radius.

To represent the dry component of the particle mobility, we introduce the dry tensor
\begin{align}
\mobility^\mathrm{dry} (a)=\frac{1}{6 \pi \eta a } \bs{I}.
\end{align}
The velocity of particle 1 has two contributions: that directly induced by the Coulomb force,  Eq.~(\ref{eq:coulomb}), 
\begin{align}
\bs{V}_1^\mathrm{dir}&=\left( \bs{\mathcal{R}} (0,\radius_w) +  \mobility^\mathrm{dry} (\radius_d) \right) \force^\mathrm{C}_{12}\nonumber\\
&= \bs{\mathcal{R}}(0,\radius_t) \force^\mathrm{C}_{12} = \mobility^\mathrm{dry} (\radius_t) \force^\mathrm{C}_{12}.\label{v1dir}
\end{align}
and the disturbance in the fluid induced by particle 2, which is subject to the opposite force,
\begin{align}
\bs{V}_1^\mathrm{dst}= -\bs{\mathcal{R}} (\position_{21},a_w) \force^\mathrm{C}_{12},\label{v1dst}
\end{align}
with the total velocity of particle 1 (parallel to the vector $\position_{12}$) being
\begin{align}
\bs{V}_1 = \bs{V}_1^\mathrm{dir} + \bs{V}_1^\mathrm{dst}.\label{v1tot}
\end{align}

\begin{figure}[h!]
   \centering
    \includegraphics[width=0.47\textwidth]{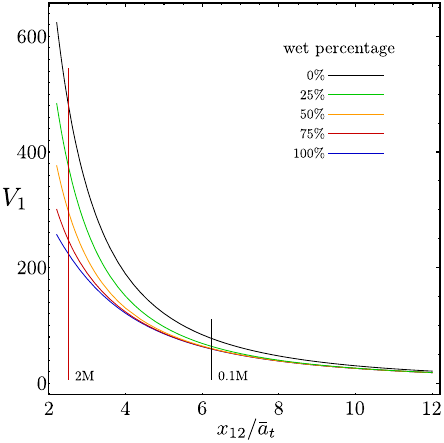}
 \caption{Particle velocity due to electrostatic and hydrodynamic interactions between two ions, as a function of separation. The velocity $V_1$ is calculated using Eqs.~(\ref{v1dir})-(\ref{v1tot}), using a radius averaged from species $A$ and $B$, i.e., $\bar{a}_t=(a_t^A+a_t^B)/2$, $\bar{a}_w=(a_w^A+a_w^B)/2$, $\bar{a}_d=(a_d^A+a_d^B)/2$. A range of wet percentages are shown, the corresponding particle radii can be found with Eqs.~(\ref{radsum}) and (\ref{diffSum}) and the radii quoted at the start of Section~\ref{sec:Results}. The measured average closest ion separations corresponding to concentrations 0.1~M and 2~M are indicated with the black and red vertical lines.}
\label{fig:wetdry1}
\end{figure}

In Fig.~\ref{fig:wetdry1} we plot $V_1$ as a function of particle separation. 
Five curves are shown, ranging from 0\% to 100\% wet, with increments of 25\%.
Also marked, with vertical lines, are the measured average nearest particle separations for two concentrations, 0.1 M (black) and 2 M (red). 
At 0.1~M varying the wetness from 50-100\% produces only a 2\% change in particle velocity. At 2~M the same range produces a change in velocity of 33\%, suggesting a higher wetness percentage is necessary to accurately capture the solution at higher molarities, as expected.

\begin{figure}[h!]
 
   \centering
    \includegraphics[width=0.47\textwidth]{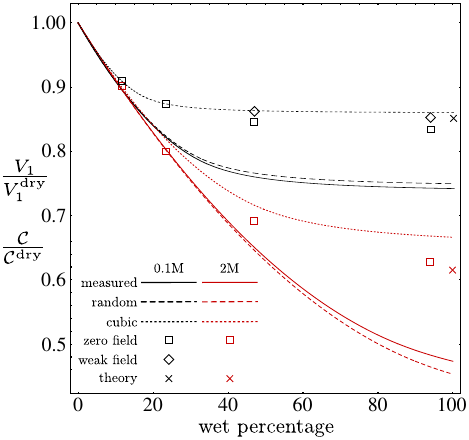}
 \caption{Normalized particle velocity due to Eq.~(\ref{v1tot}) (lines), and normalized conductivity simulation measurements from Table~\ref{condTable} (symbols), as a function of wet/dry ratio of species A. All values are normalized by the 0\% wet value. Black signifies 0.1~M, red 2~M. The curves indicate the particle velocity, calculated at the average nearest particle separation measured from simulation (solid), and corresponding to random (dashed), and cubic lattice (dotted) configurations. Squares are zero-field conductivity measurements using the Einstein-Helfand formula (Eq.~(\ref{eq:EHCond})), diamonds indicate weak-field conductivity measurements using the center of charge displacement (Eq.~(\ref{eq:simpleCond})), and crosses are the prediction of DHO theory (Eq.~(\ref{eq:FullDHO})). }
\label{fig:wetdry2}
\end{figure}

In Fig.~\ref{fig:wetdry2} particle velocity is shown as a function of wet percentage, for a range of ion separations using the same two ion system; the curves are normalized by the value at 0\% wet. In addition to the measured nearest ion separation (solid lines), for comparison we have also indicated the average nearest particle distance obtained by assuming the particles are in a random configuration (dashed lines), and in a cubic lattice (dotted lines). 
For randomly distributed particles the average nearest particle distance is \cite{feller2008introduction}
\begin{align}
    \absposition_{12} = \Gamma({\textstyle \frac43}) ({\textstyle \frac43} \pi n)^{-1/3} \approx 0.55 ~n^{-1/3},
\end{align}
where $\Gamma(x)$ is the gamma function.
Assuming a cubic lattice arrangement gives
\begin{align}
\absposition_{12}=n^{-1/3}.
\end{align}
Also shown are the conductivity simulation measurements from Table~\ref{condTable}, again normalized by the 0\% wet value. For both 0.1~M and 2~M the conductivity measurements follow a similar trend to the particle velocity, suggesting that the simple particle velocity calculation from Eqs.~(\ref{v1dir})-(\ref{v1tot}) is a useful indicator of the lower bound of wet diffusion necessary to accurately simulate a given molar concentration. For example, the change of 2\% in the particle velocity for the 0.1~M case corresponds to a 1\% change in conductivity for the two highest wetness ratios, while the 33\% change in velocity for the 2~M case corresponds to a change in conductivity of 10\%.

\section{Conclusions}\label{sec:Conclusion}
In this article we have presented
a new approach for the simulation of electrolytes at the mesoscale. The method extends the FIB method to include electrostatic interactions, and  incorporates a ``dry diffusion'' process to correct for under-resolution of the hydrodynamics. The dry diffusion term allows the method to be applied to systems of electrolytes where application of the standard FIB approach would require an excessively fine grid to resolve particles of nanoscopic size. 
We note that our approach is similar to the Stochastic Eulerian Lagrangian Method \cite{atzberger2011, atzberger2007stochastic, plunkett2014spatially} and the stochastic Force Coupling Method \cite{maxey2001localized, lomholt2003force, keaveny2014fluctuating, delmotte2015simulating}; the relationship of each of these to the FIB method is discussed in Ref.~\citenum{delong2014brownian}. In principle, some of the techniques outlined in this article could also be applied in the context of these distinct numerical approaches.

In sections \ref{subsec:GofR} and \ref{subsec:Conductivity}, we have demonstrated that the \SHIBA method accurately reproduces several important properties of electrolytes. In Sec.~\ref{subsec:wetdry}, we have demonstrated that the dry diffusion approach can effectively replace resolving the near-field hydrodynamics at scales smaller than the typical ion-ion distance. At 0.1 M, reducing the wetness to approximately 50\% produces only a 2\% change in the conductivity, but at 2 M a 10\% change is observed. Dry diffusion can therefore be used to reduce the numerical resolution needed to simulate an electrolyte, but this approach becomes less effective as the ion concentration is increased.

Our results clearly demonstrate that, for denser electrolyte solutions, near-field hydrodynamic interactions contribute to macroscopic transport properties such as conductivity. As discussed in great detail in Ref.~\citenum{donev_mesoscopic_diffusion:2014}, wet diffusion, which is diffusion due to advection by a fluctuating velocity field, and standard dry diffusion, are very different both physically and mathematically. This difference cannot be seen for a single particle since ultimately only the
 total (effective or renormalized) diffusion coefficient matters; it is worth recalling that in Ref.~\citenum{espanol2015coupling} only a single nano-particle is analyzed. The difference between wet and dry diffusion, however, manifests itself in collective properties such as the spectrum of nonequilibrium concentrations \cite{donev_mesoscopic_diffusion:2014}. However, experimentally distinguishing between wet and dry diffusion based on this spectrum requires examining length scales below light-scattering reach, and, at the scales observable in typical experiments, one only sees a total (effective or renormalized) diffusion coefficient. Our work shows that electrical conductivity, which is easily measured to high accuracy, is a sensitive probe that is affected by the hydrodynamic interactions (equivalently, the spectrum of the fluctuating velocity field~\cite{donev_mesoscopic_diffusion:2014}) at distances comparable to the typical ion-ion spacing. In the future, it is important to perform more detailed comparisons between simulations for different degrees of ``wetness'' and results for denser electrolytes obtained using experiments or large-scale molecular dynamics simulations. This will shed light on how good of a model the Rotne-Prager-Yamakawa tensor is for solvated ions.

A key advantage of the FIB method is the ability to handle other types of boundary conditions, notably confinement by no-slip walls. As the discussion of \SHIBA in this context requires considerable exposition, of both details of the algorithm and the physics that arises, we defer this to a future publication. However, as previously noted, we give some brief comments in Appendix~\ref{sec:boundaries}. It is perhaps more relevant to contrast the numerical method used in \SHIBA to the state-of-the-art Positively Split Ewald (PSE) method for Brownian Dynamics with hydrodynamic interactions developed in Ref.~\citenum{SpectralRPY}, by using some ideas from spectral Ewald methods for electrostatics \cite{SpectralEwald_Electrostatics} combined with the idea of using fluctuating hydrodynamics to generate the far-field Brownian displacements. The PSE method includes a spectrally-accurate implementation of the Hasimoto-Ewald splitting used in GGEM \cite{hernandez2007fast}; this splitting allows one to choose the grid size \emph{arbitrarily} by using the same idea of local near-field corrections that we used here for the Poisson equation. However, at present, the PSE approach relies on Fourier-space decompositions and using the FFT algorithm, and therefore only apply to triply-periodic domains. It remains a future challenge to find a way to incorporate near-field hydrodynamic corrections to correct for under-resolution of the hydrodynamics by the Stokes solver used in FIB and \SHIBAe, while also computing Brownian motion without expensive iterations.

In future work we intend to apply \SHIBA to strong electrolyte solutions that are acids (e.g., \ch{H Cl}) or bases (e.g., \ch{Na OH}) instead of salts. Also of interest are solutions in which the Wien effect is more pronounced, such as weak electrolyte acids and bases (e.g., acetic acid, ammonium hydroxide) and strong electrolyte salts that are not 1:1 (e.g., \ch{MgCl2}). Furthermore, recent work suggests that non-aqueous solvents with low permittivity (e.g., benzene) exhibit unusual Wien effects \cite{WienPRL2020}.

Here we have described the \SHIBA approach for simple mono-atomic, 1:1 electrolytes, where each ion is described by a single IB kernel. In principle, multiple kernels may be combined to form complex molecules such as charged polymers or DNA strands. This approach has already been tested in the context of BD, for example in Refs.~\citenum{Stoch_DNA} and \citenum{confine_DNA}, and for electrically neutral systems using FIB, for example in Refs.~\citenum{sprinkle2019brownian,sprinkle_2017,balboa2017hydrodynamics}. Future work may also include the extension of \SHIBA to encompass these cases. Further, there is the possibility of simulating large molecules, such as DNA, in a solution represented as a continuum that includes smaller ionic species. This was done previously using a deterministic solution in Ref.~\citenum{hernandez2015self}, in this case the solution would be modeled using the fluctuating hydrodynamic approaches outlined in Refs.~\citenum{peraud2016low,donev2019fluctuating,donev2019fluctuating2}.

\section*{Acknowledgements}

The authors thank Changho Kim for discussions regarding electrolyte chemistry, and Charles Peskin and Jonathan Goodman for discussions regarding the IB-P3M method.
This work was supported by the U.S.~Department of Energy, Office of Science, Office of Advanced Scientific Computing Research, Applied Mathematics Program under contract No.~DE-AC02-05CH11231.
This research used resources of the National Energy Research Scientific Computing Center, a DOE Office of Science User Facility supported by the Office of Science of the U.S. Department of Energy under Contract No.~DE-AC02-05CH11231.
A. Donev was supported in part by the Division of Chemical, Bioengineering, Environmental and Transport Systems of the National Science Foundation under award CBET-1804940.

\appendix

\begin{appendix}

\section{\label{sec:IB-P3M}Accuracy of immersed-boundary P3M solver}

In this appendix we examine the accuracy of the immersed-boundary P3M method described in Sections~\ref{subsec:Electrostatics} and \ref{subsec:ElectroSolver}. In particular, we focus on the degree of translational and rotational invariance of the electrostatic forces computed by the method for a pair of ions of equal and opposite charge $q_i=-q_j$ at a distance $r$ from each other, using a uniform regular grid of spacing $\Delta r$. The periodic box size $L\gg r$ so that periodic effects can be neglected.

We include in our tests both the 4-point (see Eq.~\eqref{peskin1d}) and 6-point \cite{New6ptKernel} Peskin kernels, given by
\begin{align}
\delta^\mathrm{Pe}(\zeta_k)= 
\hspace{-1mm}\begin{cases} 
       0, &\hspace{7mm}  \zeta_k \leq -3 \\
     \varphi(\zeta_k), &  \hspace{-2mm} -3 <  \zeta_k  \leq -2 \vspace{2mm} \\
      -3\varphi(\zeta_k) - \frac{\displaystyle 1}{\displaystyle 16} + \frac{\displaystyle K + \zeta_k^2}{\displaystyle 8}+ \frac{\displaystyle (3K-1) \zeta_k}{\displaystyle 12} +\frac{\zeta_k^3}{\displaystyle 12}, & \hspace{-2mm} -2 <\zeta_k \leq -1  \vspace{2mm}\\
      2\varphi(\zeta_k) + \frac{\displaystyle 1}{\displaystyle 4} + \frac{\displaystyle (4-3K)\zeta_k}{\displaystyle 6} - \frac{\displaystyle \zeta_k^3}{\displaystyle 6}, & 
      \hspace{-2mm} -1 <\zeta_k \leq 0  \vspace{2mm}\\
           2\varphi(\zeta_k) + \frac{\displaystyle 5}{\displaystyle 8} - \frac{\displaystyle K + \zeta_k^2}{\displaystyle 4}, & 
      \hspace{1mm} 0 <\zeta_k \leq 1  \vspace{2mm}\\
      -3\varphi(\zeta_k) + \frac{\displaystyle 1}{\displaystyle 4} - \frac{\displaystyle (4-3K)\zeta_k}{\displaystyle 6}+ \frac{\displaystyle \zeta_k^3}{\displaystyle 6}, & 
      \hspace{1mm} 1 <\zeta_k \leq 2  \vspace{2mm}\\
       \varphi(\zeta_k) - \frac{\displaystyle 1}{\displaystyle 16} + \frac{\displaystyle K+\zeta_k^2}{\displaystyle 8}- \frac{\displaystyle (3K-1)\zeta_k}{\displaystyle 12}- \frac{\displaystyle \zeta_k^3}{\displaystyle 12}, & 
      \hspace{1mm} 2 <\zeta_k \leq 3  \vspace{2mm}\\
       0, &\hspace{1mm}  3<\zeta_k 
   \end{cases},\label{peskin6pt}
\end{align}
where
\begin{align*}
&\varphi(\zeta_k)=\left(-\vartheta(\zeta_k) + \rm{sgn}\left(\frac{\displaystyle 3}{\displaystyle 2}-K\right)\sqrt{\vartheta^2(\zeta_k)-112\varsigma(\zeta) }\right)/56,\\
&\vartheta(\zeta_k) = 9/4-(3/2)(K+\zeta_k^2)\zeta_k+(22/3-7K)\zeta_k-(7/3)\zeta_k^3,\\
&\varsigma(\zeta_k) = -(11/32)\zeta_k^2 + (3/32)(2K+\zeta_k^2)\zeta_k^2+(1/72)\left((3 K-1)\zeta_k + \zeta_k^3\right)^2 +(1/18)\left( (4-3K)\zeta_k-\zeta_k^3 \right)^2,\\
&K=59/60-\sqrt{29}/20.
\end{align*}
In order to compare to alternative methods, these tests were performed in Matlab using FFTs instead of multigrid to solve the (discrete) Poisson equation. In addition to the 2nd-order electrostatic solver based on the 7-point Laplacian $(\graddiscrete^2)^c$ and centered difference $\graddiscrete^{c}$, we also include, for comparison, results obtained by using a fourth-order compact isotropic 24-point Laplacian in Eq.~\eqref{eq:poissondisc} and a 4-point centered difference in Eq.~\eqref{eq:fielddisc}. 

In addition to the fourth-order finite-difference Poisson solver, we also implement a spectral Fourier solver, as used in the Spectral Ewald (SE) method \cite{SpectralEwald_Electrostatics}. As explained in Section~\ref{subsec:Electrostatics}, the SE approach (and essentially all other Ewald methods) uses a Gaussian of standard deviation $\sigma$ as the kernel $\electrokernel$ in Eqs.~\eqref{kernel3} and \eqref{kernel4}. This allows one to compute the electrostatic force between the two Gaussian charge clouds analytically for an unbounded domain,\footnote{Note that the periodic boundaries are implemented by using a Fourier series instead of a Fourier sum.} giving the required local correction
\begin{equation}
    F^\mathrm{LC}_{\text{Gaussian}}= 
    \left(\frac{\Delta r}{r}\right)^2 \left(
    \text{erfc}\left(\frac{r}{2\sigma}\right) 
    + \frac{r}{ \sigma \sqrt{\pi}}\exp\left(-\frac{r^2}{4\sigma^2}\right)
    \right)
    \label{eq:coulomb_Gaussian}
\end{equation}
that decays exponentially with distance.
However, achieving 3-4 digits of accuracy with Gaussian kernels requires the support of the kernel to be on the order of 10-12 grid points in each dimension, which is expensive. Therefore, to make a more fair comparison with the 6-point Peskin kernel, here we couple the spectral Poisson solver with a recently-developed ``exponential of a semicircle'' (ES) kernel \cite{FINUFFT_Barnett} with width $w=6$; we optimized the value of the parameter $\beta=1.8w$ for translational and rotational invariance.\footnote{We point out again that this approach, while similar to the Spectral Ewald method \cite{SpectralEwald_Electrostatics} based on the non-uniform FFT \cite{FINUFFT_Barnett} using the ES kernel, is different because we do use the same kernel to both smear the charges and to communicate to the grid, rather than using two different kernel functions (Gaussian for smearing and ES for grid transfers in the NUFFT).}

\begin{figure}[h!]
 
  \centering
    \includegraphics[width=0.47\textwidth]{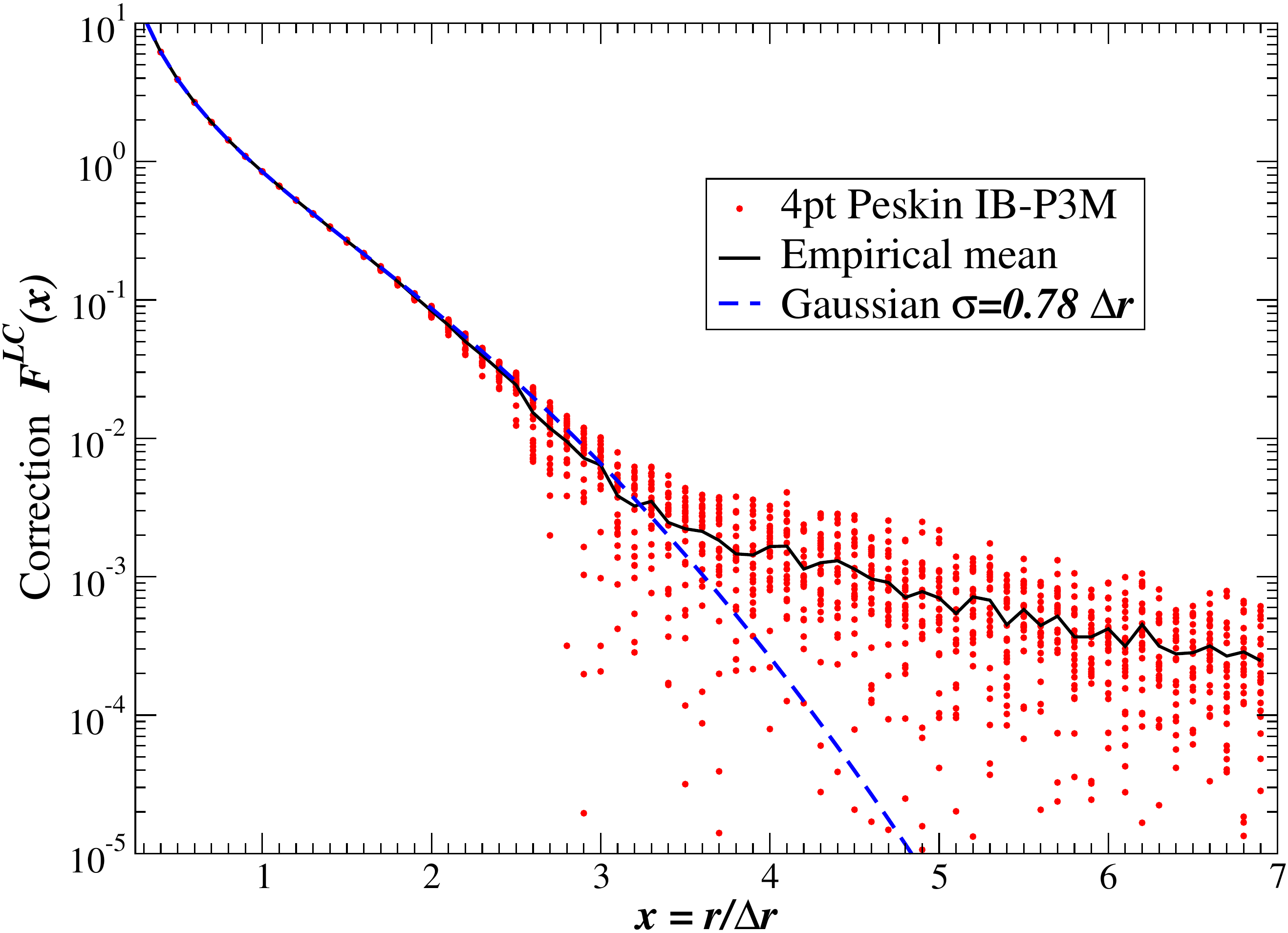}
    \includegraphics[width=0.47\textwidth]{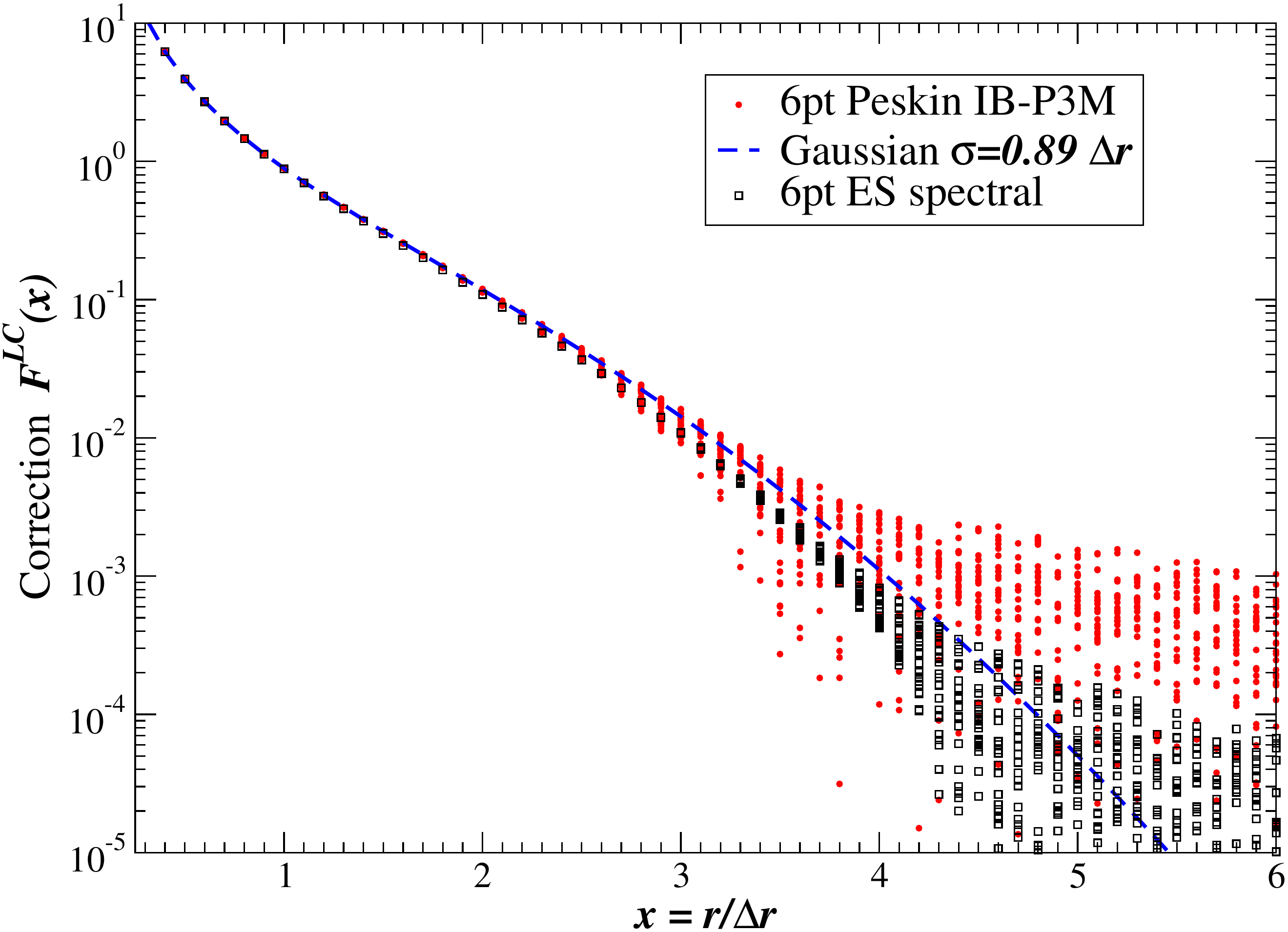}
\caption{\label{fig:IB-P3M-scatter}Local correction function $F^\mathrm{LC}(x=r/\Delta r)$ computed from the radial component of the near-field correction to the electrostatic force, $\force_{ij}^\mathrm{LC}$, for the Peskin 4pt kernel (top panel) and the Peskin 6pt kernel (bottom panel). Red circles are samples for different positions and orientations of the pair of ions, while dashed blue lines indicate the continuum theoretical near-field correction $F^\mathrm{LC}_{\text{Gaussian}}$ for Gaussian charges of standard deviation $\sigma$ indicated in the legend (see Eq.~\eqref{eq:coulomb_Gaussian}). The black solid line in the top panel indicates the empirical mean used to tabulate $F^\mathrm{LC}$ in our method. The bottom panel includes results (black squares) for a spectral Poisson solver with a 6-point exponential of a semicircle (ES) kernel \cite{FINUFFT_Barnett} with parameter $\beta=1.8\times6$.}
\end{figure}

Fig.~\ref{fig:IB-P3M-scatter} shows numerical data for the radial component of the near-field correction to the electrostatic force $\force_{ij}^\mathrm{LC}$ for several kernels and Poisson solvers, expressed in terms of the dimensionless function $F^\mathrm{LC}(x=r/\Delta r)$ appearing in Eq.~\eqref{eq:F_LC}. To quantify the loss of translational and rotational invariance, we randomly generate the position of many pairs of particles at a distance $r$ from each other. Note that if $\force_{ij}^\mathrm{LC}$ were purely radial and its magnitude were purely a function of $r$, the local-correction approach used here could completely correct the electrostatic force computed by the discrete solver to match the Coulomb force between point charges. Therefore, the main cause of inaccuracy is the loss of translational and rotational invariance.
In Fig.~\ref{fig:IB-P3M-scatter} we also show the correction $F^\mathrm{LC}(x=r/\Delta r)$ for a pair of Gaussian charge clouds given in Eq.~\eqref{eq:coulomb_Gaussian}. The value of the Gaussian width $\sigma$ was fitted to match the numerical results over the range $0<x<3$, giving $\sigma\approx0.78\Delta r$ for the 4-point and $\sigma\approx0.89\Delta r$ for the 6-point Peskin kernels with the 2nd order Poisson solver. This can be thought of as an effective width of the charges similar to the effective hydrodynamic radius $a_w$ for Stokes flow.

The spread in the symbols in Fig.~\ref{fig:IB-P3M-scatter} shows the loss of translational and rotational invariance. We see in the top panel that the mean correction force decays algebraically rather than exponentially as it does for Gaussian charges. We also see that the 6pt kernel improves the invariance, especially at short distances. Using a fourth-order isotropic Poisson solver further improves the invariance for all distances (data not shown in Fig.~\ref{fig:IB-P3M-scatter}), as does using a spectral Poisson solver with the 6-point ES kernel, as shown in the bottom panel of Fig.~\ref{fig:IB-P3M-scatter}. For the 4-point kernel, we see in the top panel of the figure that truncating the local corrections for pairs of particles further than 3-grid points apart, $\psi=3\Delta r$, makes the error due to loss of invariance dominate the error due to truncation of $F^\mathrm{LC}(x=r/\Delta r)$, and therefore there is no point in further increasing the cutoff distance $\psi$; similarly, for the 6-point kernel $\psi=4\Delta r$ is suitable.

\begin{figure}[h!]
\begin{minipage}{0.47\textwidth}
    \includegraphics[width=1\textwidth]{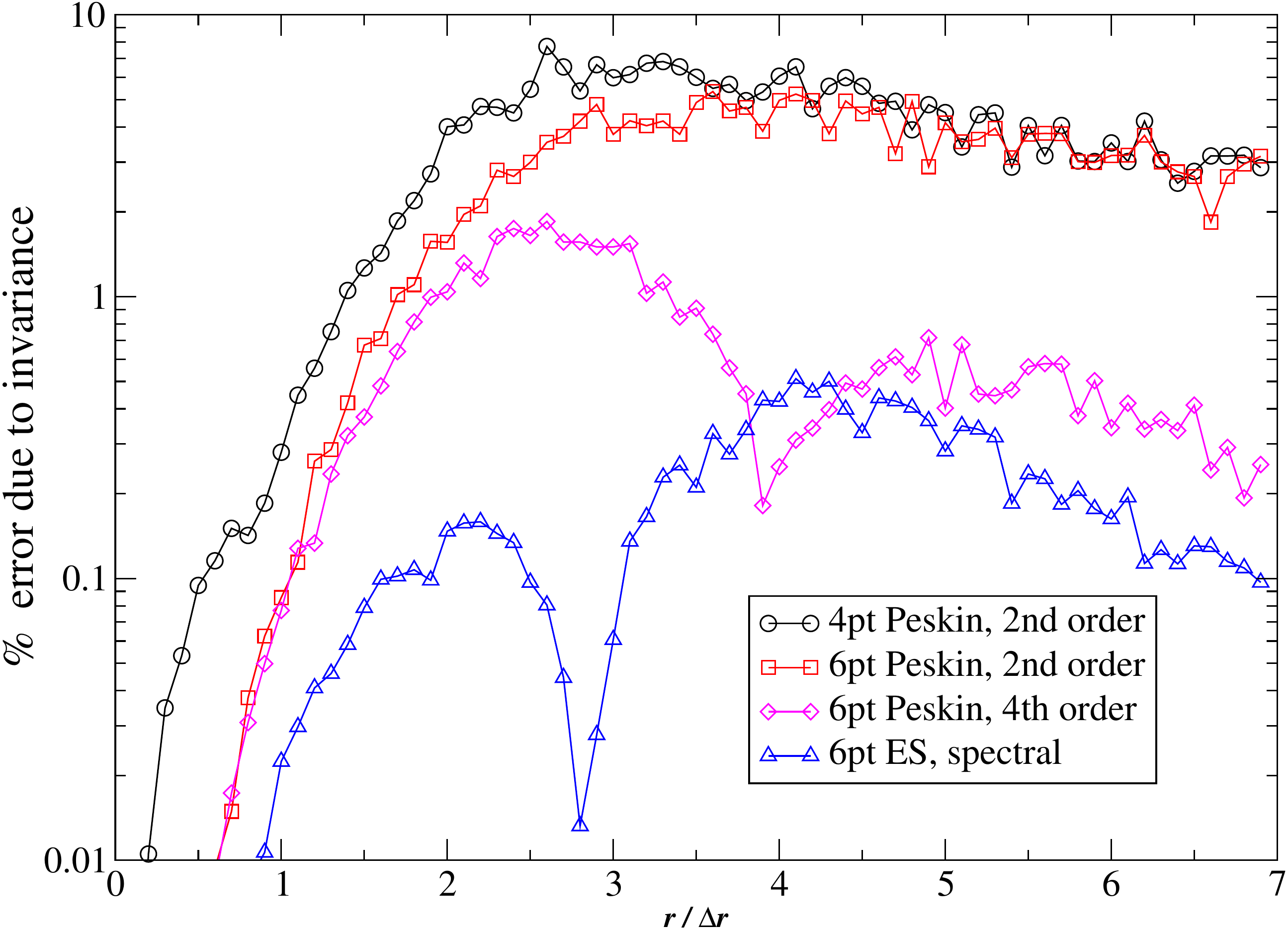}
\end{minipage}
\begin{minipage}{0.5\textwidth}
   \vspace{3mm}    
\includegraphics[width=1\textwidth]{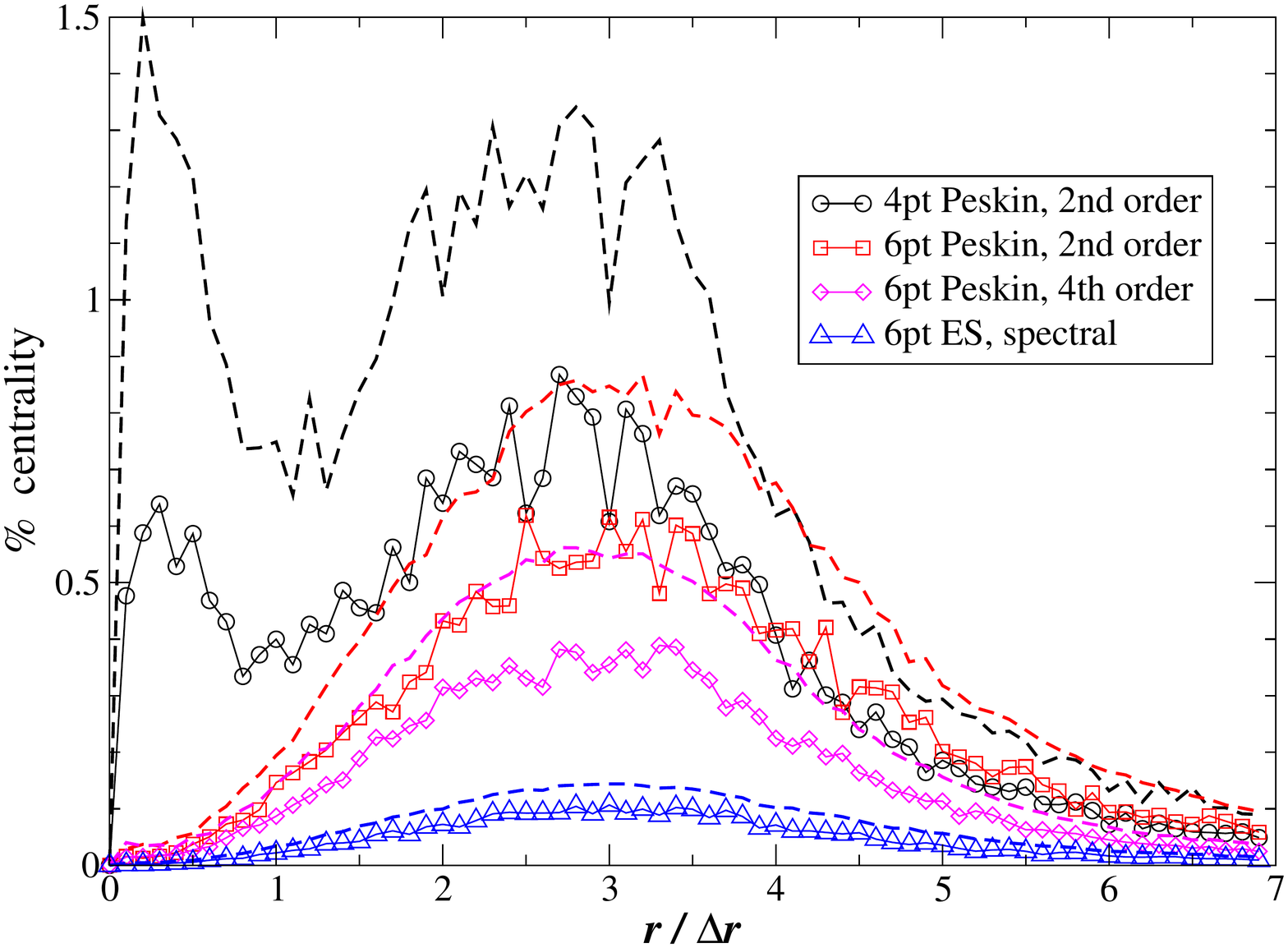}
\end{minipage}
\caption{\label{fig:IB-P3M-invariance}Percent error due to loss of translational and rotational invariance in the immersed-boundary P3M method based on data shown in Fig.~\ref{fig:IB-P3M-scatter}. In addition to the method used in this work, we include for comparison the improvement gained by using a fourth-order isotropic Poisson solver, as well as a spectral Poisson solver with a 6-point ES kernel. (Left) Percent error in the radial component of the total electrostatic force due to loss of invariance. (Right) Magnitude of the non-radial component of the IB force $F^\mathrm{P}$ expressed as percent of the radial component. The symbols show the empirical mean over samples of pairs of points, while the dashed lines show the maximum value over the samples.
}

\end{figure}

We further quantify the amount of translational and rotational variance in Fig.~\ref{fig:IB-P3M-invariance}. The top panel shows the spread in the symbols in Fig.~\ref{fig:IB-P3M-scatter} for each distance $r$, measured as two standard deviations of the samples. We normalize the spread by $1/x^2$ to obtain a measure of the invariance of the radial component of the electrostatic force relative to the true Coulomb potential, and express the error in percent; note that this is the error in the final answer for the electrostatic force between two ions. The bottom panel shows the magnitude of the non-radial component of the electrostatic force computed by the Poisson solver, $\force_{ij}^\mathrm{P}$, relative to it's radial component, again expressed as percent error.
The top panel of Fig.~\ref{fig:IB-P3M-invariance} shows that the second-order Poisson solver is translationally and rotationally invariant to about 8\% for the 4-point kernel and about 6\% for the 6-point Peskin kernel, with the 6-point kernel being notably more accurate at short distances. The fact that errors become comparable for large distances regardless of the width of the kernel suggests that for large distances the error is dominated by the anisotropy of the discrete Green's function for the standard 7-point Laplacian.
Switching to a fourth-order Laplacian drops the maximum error down to 2\% for the 6-point kernel, while the spectral solver with the 6-point ES kernel drops the error below 0.5\%. The bottom panel of the figure shows that all combinations of kernels and solvers studied here achieve a central electrostatic force to within 1.5\%, which is remarkably good.

\section{\label{sec:IB-P3M-Tables}P3M correction tables}
The numerical results in the main body of this work were computed using a four-point Peskin kernel \cite{peskin2002}, additional analysis in Appendix~\ref{sec:IB-P3M} uses a six-point kernel \cite{New6ptKernel}. As per Eqs.~(\ref{eq:LC}) and (\ref{eq:F_LC}), the P3M procedure uses pre-computed tables to apply a near field correction to electrostatic forces. The values for each kernel are given below in Table~\ref{peskinTable}. Note that for values of $F^\text{P}$ between the tabulated positions linear interpolation was used.
\begin{table*}[htb]
\begin{center}
\begin{tabular}{|c|c|c|c|c|c|c|c|c|c|c|c|c|c|c|}
\hline
&four-point  & six-point &&&four-point  & six-point &&&four-point  & six-point&&& six-point\\
\hline
$\frac{\displaystyle\absposition_{ij}}{\displaystyle\Delta\absspatial}$ &$F^\mathrm{P}\left(\frac{\displaystyle\absposition_{ij}}{\displaystyle\Delta\absspatial}\right)$ &$F^\mathrm{P}\left(\frac{\displaystyle\absposition_{ij}}{\displaystyle\Delta\absspatial}\right)$&&$\frac{\displaystyle\absposition_{ij}}{\displaystyle\Delta\absspatial}$ &$F^\mathrm{P}\left(\frac{\displaystyle\absposition_{ij}}{\displaystyle\Delta\absspatial}\right)$ &$F^\mathrm{P}\left(\frac{\displaystyle\absposition_{ij}}{\displaystyle\Delta\absspatial}\right)$&&$\frac{\displaystyle\absposition_{ij}}{\displaystyle\Delta\absspatial}$ &$F^\mathrm{P}\left(\frac{\displaystyle\absposition_{ij}}{\displaystyle\Delta\absspatial}\right)$ &$F^\mathrm{P}\left(\frac{\displaystyle\absposition_{ij}}{\displaystyle\Delta\absspatial}\right)$&&$\frac{\displaystyle\absposition_{ij}}{\displaystyle\Delta\absspatial}$ &$F^\mathrm{P}\left(\frac{\displaystyle\absposition_{ij}}{\displaystyle\Delta\absspatial}\right)$  \\
\hline
&&&&&&&& &&&&&\\
0&0&0&&2.0&0.165755&0.133308&&4.0&0.0627171&0.0616157&&6.0&0.0275866\\
0.1&0.0191993&0.0130335&&2.1&0.162663&0.130995&&4.1&0.0591556&0.0586423&&6.1&0.0267607\\
0.2&0.0384104&0.0259347&&2.2&0.156037&0.129343&&4.2&0.0561682&0.0561644&&6.2&0.0259347\\
0.3&0.0561438&0.0384891&&2.3&0.148626&0.126205&&4.3&0.0539003&0.0535214&&6.3&0.0251088\\
0.4&0.0745141&0.0507132&&2.4&0.144648&0.122075&&4.4&0.0511791&0.0513739&&6.4&0.0242829\\
0.5&0.0914394&0.0624416&&2.5&0.138100&0.118276&&4.5&0.0492405&0.0487309&&6.5&0.0236221\\
0.6&0.106869&0.0735093&&2.6&0.131473&0.115137&&4.6&0.0470903&0.0469138&&6.6&0.0229613\\
0.7&0.121719&0.0837511&&2.7&0.12615&0.110677&&4.7&0.0453286&0.0450967&&6.7&0.0221354\\
0.8&0.134211&0.0931669&&2.8&0.120057&0.106547&&4.8&0.043644&0.0431145&&6.8&0.0214746\\
0.9&0.145262&0.101922&&2.9&0.112581&0.102748&&4.9&0.0417381&0.0414626&&6.9&0.0209791\\
1.0&0.154958&0.109521&&3.0&0.106808&0.0979574&&5.0&0.0399759&&&&\\
1.1&0.162921&0.116293&&3.1&0.100863&0.0936624&&5.1&0.038324&&&&\\
1.2&0.168841&0.121745&&3.2&0.0947845&0.0898631&&5.2&0.0366721&&&&\\
1.3&0.173264&0.126535&&3.3&0.0905315&0.0860637&&5.3&0.0355157&&&&\\
1.4&0.176417&0.130335&&3.4&0.0856055&0.0816036&&5.4&0.034029&&&&\\
1.5&0.17854&0.132978&&3.5&0.0818209&0.0782998&&5.5&0.0328727&&&&\\
1.6&0.177695&0.134134&&3.6&0.0768459&0.0745005&&5.6&0.0317164&&&&\\
1.7&0.175969&0.135621&&3.7&0.0728331&0.0707011&&5.7&0.0307252&&&&\\
1.8&0.174664&0.13529&&3.8&0.0694836&0.0680581&&5.8&0.0295689&&&&\\
1.9&0.170830&0.134629&&3.9&0.0650267&0.0645891&&5.9&0.0285778&&&&\\
  \hline
\end{tabular}
\end{center}
\caption{Tabulated values for P3M correction using four-point and six-point Peskin kernels, as described in Sec.~\ref{subsec:Electrostatics}. Note that data has been shown beyond the ideal cutoff range, $\psi$; see Appendix~\ref{sec:IB-P3M}.}\label{peskinTable}
\end{table*}

\FloatBarrier
\section{\label{sec:boundaries}Comments on non-periodic boundaries}
As mentioned in sections \ref{sec:Introduction} and \ref{sec:Conclusion}, one advantage of the DISCOS method is the ability to handle non-periodic, or confining, boundary conditions with relative ease. As the techniques used to do so require some exposition, and a range of case studies are required to validate these techniques, we defer a complete description to a future publication. However here we give a brief outline of some of the issues.

There are two major considerations when incorporating confining boundary conditions: (i) boundary conditions for the P3M electrostatic interactions, and (ii) boundary interactions for the dry component of the solvent. We now describe each of these in turn, then give several example simulations. Non-periodic boundary conditions can easily be incorporated in a finite volume discretization of Poisson's equation. The short range corrections adjacent to boundaries are included using image particles \cite{jackson2007classical}. In this case $\force_{ij}^\mathrm{LC}$,
Eq.~(\ref{eq:LC}), is trivial to calculate using the location of the image particles.

For the wet component of the hydrodynamic interactions, no additional steps are required to add confining boundaries; these are trivially incorporated into the GMRES method. In particular, extensive discussion of no slip walls in the context of the FIB method is given in Ref.~\citenum{delong2014brownian}.  An adjustment to the dry component of the particle mobility is required to account for nearby boundaries. For some simple geometries there are analytic solutions, e.g. a formula is available for a single infinite plane. For general geometries, the modification to the dry mobility must be pre-computed by solving Stokes' equation on a grid corresponding to a fully wet simulation.

Below in Fig.~\ref{fig:boundary} we give two examples of simulations performed using the above approaches. In each case an infinite slit channel is used, i.e.~periodic boundaries in two dimensions and confining boundaries in the third. In the first we show the distribution of particles when homogeneous Dirichlet and homogeneous Neumann boundary conditions are used for the hydrodynamic and electrostatic solutions, respectively; this corresponds closely to a no slip dielectric surface with a material of low dielectric constant on the outside of the channel. Additionally, a constant potential drop has been applied across the channel. The DISCOS solution is compared to a molecular dynamics result from Ref.~\citenum{jing2015ionic}, and excellent agreement is observed. In the second example an electric field is applied parallel to the channel walls, and an inhomogeneous Neumann condition is applied to the electrical potential; this corresponds to a wall with surface charge. To ensure overall charge neutrality an imbalance of positive and negative ions is present in the channel itself, inducing an electro-osmotic flow. The resulting solvent velocity profile is compared to a deterministic version of the method given in Ref.~\citenum{peraud2016low}. Reasonable agreement is observed; the 9\% difference in the peak velocity is possibly due to the absence of thermal fluctuations in the deterministic continuum solution.

\begin{figure}[h!]

    \includegraphics[width=0.95\textwidth]{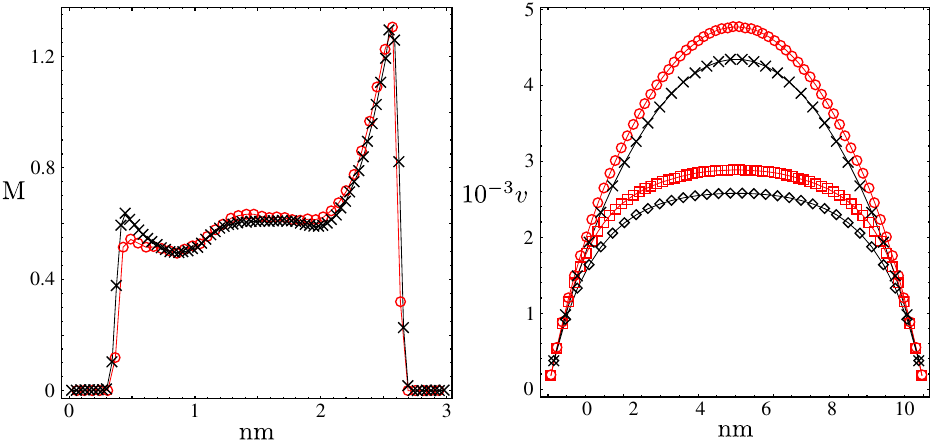}

\caption{\label{fig:boundary}Left: Comparison of cation molarity between DISCOS (red) and MD (black), for a channel with dielectric boundaries and a wall normal electric field. All simulation parameters have been selected to match those in Ref.~\citenum{jing2015ionic}. Right: Comparison of electro-osmotic flow between DISCOS and the continuum electrolyte code described in Ref.~\citenum{peraud2016low}. The red circles and black crosses represent the DISCOS and continuum fluid velocities a short time after the electric field has been applied. The red squares and black diamonds show the flow at large time when the system has reached steady state.}

\end{figure}

\end{appendix}

\bibliography{FHDX}

\end{document}